\documentclass[apj]{emulateapj}

\usepackage{graphicx}
\usepackage{multirow}
\usepackage{epsfig}
\usepackage{amsmath}
\slugcomment{Accepted for publication in AJ}
\shorttitle{Stacking Galaxy Images}
\shortauthors{Li et al.}

\newcommand\sersic{S$\acute{\rm e}$rsic }

\begin{document}

\title{Measuring Structural Parameters  Through Stacking Galaxy Images}

\author{Yubin~Li\altaffilmark{1,2}, Xian~Zhong~Zheng\altaffilmark{1}, Qiu-Sheng~Gu\altaffilmark{3,4}, Yi-Peng~Wang\altaffilmark{3,4}, Zhang~Zheng~Wen\altaffilmark{1,2}, Kexin~Guo\altaffilmark{1}, Fang~Xia~An\altaffilmark{1,2}}

\altaffiltext{1}{Purple Mountain Observatory, Chinese Academy of Sciences, 2 West Beijing Road, Nanjing 210008, China; xzzheng@pmo.ac.cn}
\altaffiltext{2}{University of Chinese Academy of Sciences, 19A Yuquan Road, Beijing 100049, China}
\altaffiltext{3}{School of Astronomy and Space Science, Nanjing University, Nanjing 210093, China}
\altaffiltext{4}{Key Laboratory of Modern Astronomy and Astrophysics, Nanjing University, Ministry of Education, Nanjing 210093, China}

\begin{abstract}
It remains challenging to detect the low surface brightness structures of faint high-$z$ galaxies, which is key to understanding the structural evolution of galaxies. The technique of image stacking allows us to measure the averaged light profile beneath the detection limit and probe the extended structure of a group of galaxies. We carry out simulations to examine the recovery of the averaged surface brightness profile through stacking model {\it HST}/ACS images of a set of galaxies as functions of \sersic index ($n$), effective radius ($R_{\rm e}$) and axis ratio ($AR$). The \sersic profile best fitting the radial profile of the stacked image is taken as the recovered profile, in comparison with the intrinsic mean profile of the model galaxies. Our results show that, in general, the structural parameters of the mean profile can be properly determined through stacking, although systematic biases need to be corrected when spreads of $R_{\rm e}$ and $AR$ are counted. We find that \sersic index is slightly overestimated and $R_{\rm e}$ is underestimated at $AR<0.5$ as the stacked image appears to be more compact due to the presence of inclined galaxies; the spread of $R_{\rm e}$ biases the stacked profile to have a higher \sersic index. We stress that the measurements of structural parameters through stacking should take these biases into account.  We estimate the biases in the recovered structural parameters from stacks of galaxies when the samples have distributions of $R_{\rm e}$, $AR$ and $n$ seen in local galaxies.

\end{abstract}

\section{INTRODUCTION} \label{sec:intro}

The evolution of galaxies are found both theoretically and observationally to correlate with stellar mass \citep[e.g.,][]{Kauffmann03,Bundy06,Dekel06,Guo08,Peng10}, yielding fundamental relationships between stellar mass and  color \citep{Baldry04}, size \citep{shen03}, metallicity \citep{Tremonti04} and star formation rate \citep{Brinchmann04} among local galaxies \citep[see][for a review]{blanton09}. These relationships evolve significantly out to high redshifts \citep[e.g.,][and references therein]{Erb06,zheng07,brammer11,wuyts10,wuyts11,Shapley11}.  Much effort has been made to characterize the structural properties of galaxies at different cosmic epochs in order to dissect different physical processes regulating galaxy evolution \citep[][]{Conselice14}.
  The  size of massive galaxies has been found to increase on average by a factor of $\sim 3 - 5$ since $z\sim$2 \citep[e.g.,][]{Trujillo06,Trujillo07,Toft07,Zirm07,vdw08,mancini10,Damjanov11,Newman12,Krogager13,Belli14,vdwel14}. \citet{vdk10} found that an extended stellar halo around massive galaxies was gradually built up over cosmic time, suggesting that accretion of satellite galaxies plays a key role in governing the size growth of the massive galaxies \citep{naab09,Oser10}.  While physical interpretations of the dramatic size evolution are still under debate \citep[e.g.,][]{Hopkins10}, the spatially-resolved brightness profile as a function of redshift turn out to be crucial to unveiling the assembly histories of galaxies \citep[e.g.,][]{Trujillo11,Hilz13}.  In particular, the brightness profiles of low-mass galaxies at high-$z$ are poorly explored.

It is technically challenging to measure the brightness profiles towards large radius for typical ($L^\ast$) and low-mass galaxies at high redshifts even with deep imaging of the {\it Hubble} Space Telescope ({\it HST}) \citep{Szomoru12}.
The size of galaxies may be underestimated if the extended structure of low surface brightness is not detected  \cite[e.g.,][]{araa94,bezanson09,naab07,naab09,mancini10}. Stacking is a powerful tool to  suppress background noise and detect fluxes beneath the detection limit for individual objects. It has been applied successfully in studies with optical \cite[e.g.,][]{zibetti04,vdw08,vdk10}, infrared \cite[e.g.,][]{zheng06,lee10,bourne12,Guo13}, and radio \cite[e.g.,][]{white07,garn09,hancock11} imaging data.
\citet{vdk10} examined the systematical effects in parameterizing the mean structure of massive galaxies via stacking ground-based images, finding that the averaged size and \sersic index can be recovered when each of the stacked galaxies is characterized by a single \sersic profile.
In practice, galaxies tend to have multiple components (e.g., bulge+disk) with different surface brightness profiles; galaxies of similar stellar masses have effective radius and/or axis ratio (or inclination angle) spanning over a range \citep{shen03, hao06, padilla08}.
Further investigation is demanded to address how the  scatter in effective radius, axis ratio and \sersic index effect on the recovered structural parameters from the stacked images and to which extent the results of stacking are accurate and robust.

 In this paper, we present the results of our simulation to characterize the dependences of the averaged structural parameters of faint galaxies derived from stacking on effective radius ($R_{\rm e}$), axis ratio ($AR$), index of \sersic profile ($n$),  and the distributions of these parameters.
We describe our methodology in Section~\ref{sec:method}. In Section~\ref{sec:results} we present the simulation results. We discuss our results and summarize them in Section~\ref{sec:disc}. We assume a cosmology with $H_0$=70\,km\,s$^{-1}$\,Mpc$^{-1}$, $\Omega_M$=0.3, and $\Omega_{\Lambda}$=0.7 throughout this paper.

\section{METHODOLOGY} \label{sec:method}

\subsection{Galaxy Models}

The existing deep optical and near-infrared imaging data from large surveys with {\it HST}, including GEMS \citep{Rix04}, COSMOS \citep{Scoville07} and CANDELS \citep{Grogin11}, provide the basis for a stacking analysis of faint high-$z$ galaxies. In particular, the {\it HST}/Advanced Camera for Surveys (ACS) imaging of  COSMOS  over  1.48\,deg$^2$ \citep{Koekemoer07} through the $F814W$ ($i$) filter allows for morphological examination for large samples of galaxies.
In our simulations, we adopt a pixel size of  0$\farcs$05 (same as {\it HST}/ACS pixel size) and ACS Point Spread Function (PSF) in the $i$ band to generate galaxy model images. A physical scale of 100\,kpc then corresponds to [280,~250,~239] pixels at $z$=[0.7,~1,~2]. A size of $351\times 351$ pixels is chosen for the model images to have the radial surface brightness profile extended to $R=50$\,kpc and have background estimation out to $R=70$\,kpc  for galaxies at $z>0.7$.

Three structural parameters are used to characterize the two-dimensional model image of a galaxy:  index of \sersic profile ($n$), effective radius ($R_{\rm e}$) and axis ratio ($AR$) . Position angle is randomly chosen between 0 and 180 degrees for $AR<1$.
The \sersic profile is described by
\begin{equation}
 I(r) = I_0\,\exp\{-b_n [(r/r_{\rm e})^{1/n} - 1]\}.
\end{equation}
The axis ratio ($AR$), defined as the ratio of minor axis $b$ over major axis $a$, measures the elongation in morphology for early-type galaxies or the inclination for late-type galaxies.
The model galaxy is centered at the centroid of the image, which is then convolved with the empirical PSF derived from {\it HST}/ACS $i$-band images using a number of isolated bright stars. The Full Width at Half Maximum (FWHM) of the PSF is $R_{\rm PSF}=0\farcs 11$ or 2.2\,pixels.
The total brightness of the model image is scaled to match the total flux in analog-digital units (ADU) for a galaxy ranges from 24\,mag to 24.75\,mag (i.e., a factor of 3 spread in flux) in the {\it HST} images from COSMOS. We adopt a faint-end slope of $-$0.47 for early-type galaxies and $-$1.37 for late-type galaxies to determine the distribution of magnitude of the galaxies \citep{Tomczak14}, and we add photon noise and background noise to match the noise level in the {\it HST} images of  COSMOS in order to address the effect of noise.
An IDL code SIMULATE\_GALAXY.PRO\footnote{http://www.mpia.de/GEMS/fitting\_utilities/simulate\_galaxy.pro} \citep[see][for more technical details]{Baussler07} is used to create a galaxy model image at given $n$, $R_{\rm e}$ and $AR$.
We also randomly locate the center of a model galaxy within a pixel to match observations. A set of galaxy model images are shifted and aligned to the same center before stacking. Due to the noises, the measured center of a model galaxy slightly differs from the actual center. We will discuss the effects of this issue to the final conclusions in Section~\ref{sec:offset}.

\begin{figure*}
\begin{center}
\includegraphics[width=0.88\textwidth]{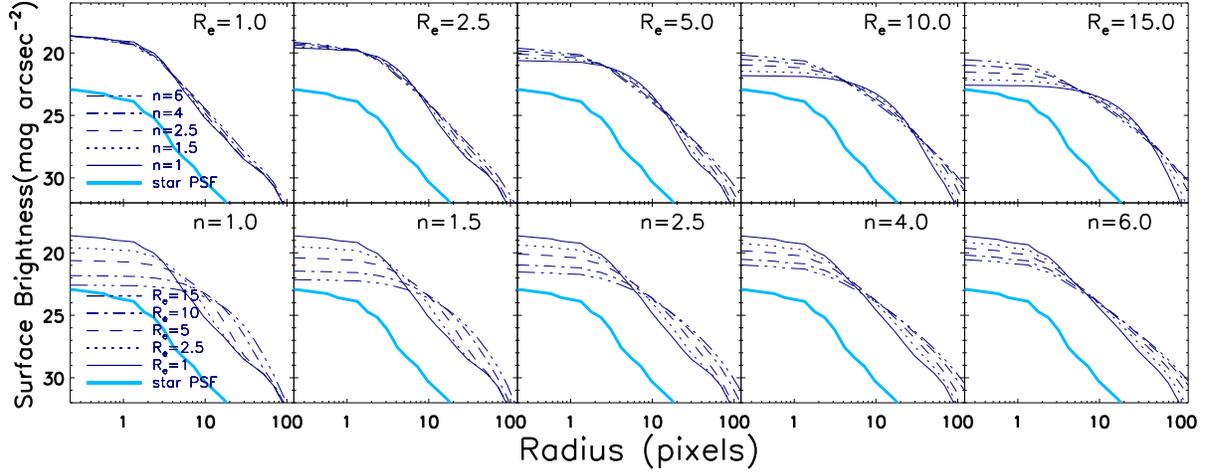}
\caption{Comparison of surface brightness profiles with different $n$ and $R_{\rm e}$.  Cyan thick lines are the PSF profile arbitrarily shifted downward for clarity.}
\label{fig:fig1}
\end{center}
\end{figure*}

\begin{figure*}
\begin{center}
\includegraphics[width=0.88\textwidth]{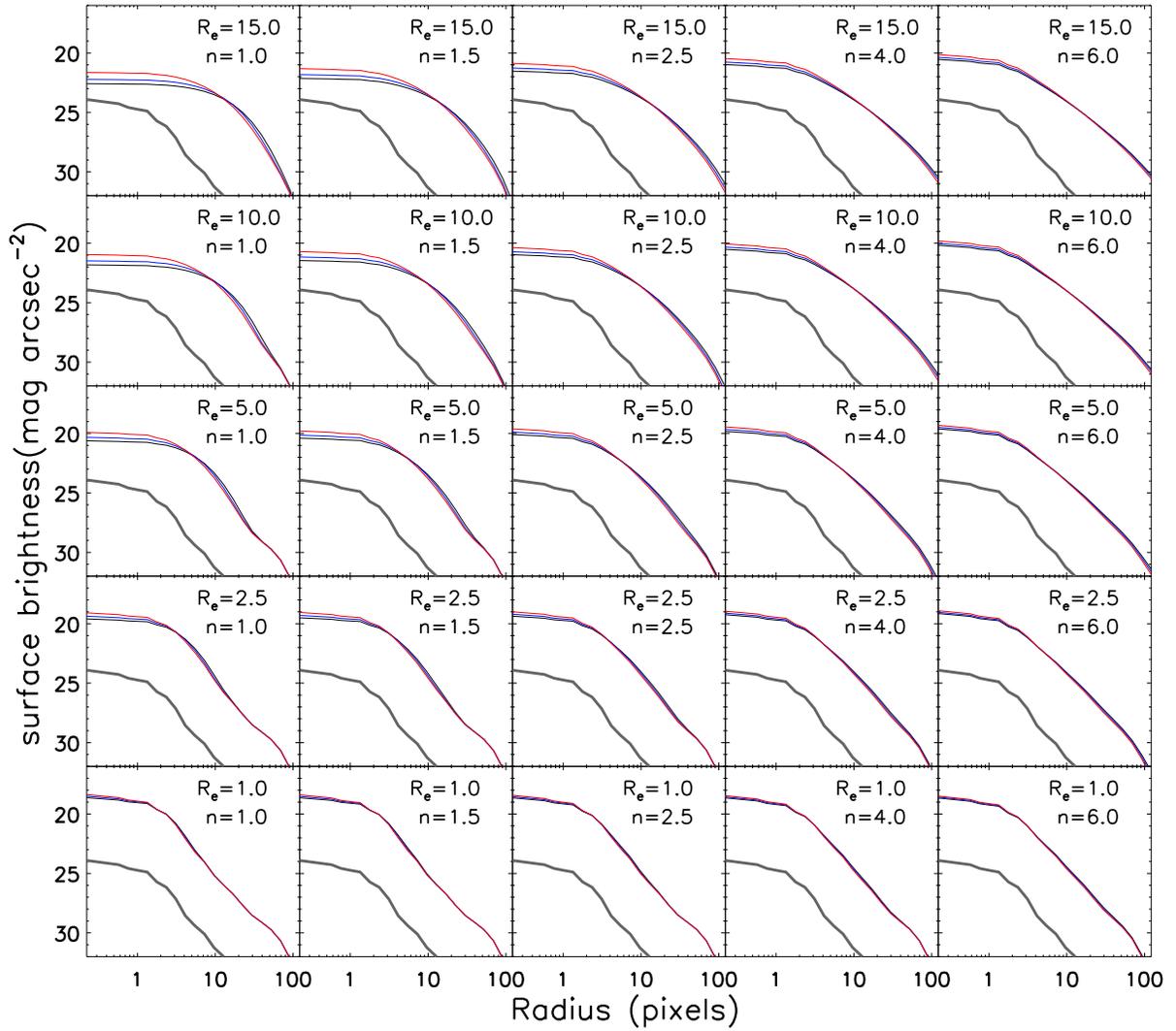}
\caption{Comparison of the surface brightness profiles extracted using circular apertures from model galaxies of given axis ratios ($AR$). Black lines show the galaxy models with $AR=1$ (face-on).  Red lines represent $AR=0.34$ and blue lines represent $AR=0.68$. Grey thick lines show the PSF profile arbitrarily shifted downward for clarity.}
\label{fig:fig2}
\end{center}
\end{figure*}

In practice, the structural parameters are barely known for individual faint high-$z$ galaxies; the stacked image is often obtained by directly co-adding the aligned galaxy images without corrections for inclination, orientation and size \citep[but see, e.g.,][]{Zibetti05}; the stacked surface brightness profile is usually derived from the stacked image using circular apertures.
We generate galaxy models with structural parameters spanning over sufficiently wide ranges: $1\le n \le 6$, $0\farcs 05 \le R_{\rm e} \le 0\farcs 75$ and $0.1 \le AR \le 1.0$.
Firstly, we examine the measurement of  surface brightness  profiles derived using circular apertures from a single model image  in relations to each of  structural parameters $n, R_{\rm e}$ and $AR$.
Secondly, we test the recovery of the mean surface brightness profile when a set of stacked galaxies have two parameters fixed and the third parameter following a certain distribution. Each set contains  700 model images, which is representative of practical cases for stacking.
We divide galaxy models into late-type ($1\le n \le 2.5$) and early-type ($2.5 < n \le 6$) because the two populations are distinct in structure \citep{blanton09}. We adopt the  log-normal distributions  of $R_{\rm e}$ for the two populations from \cite{shen03}. The  log-normal distribution of $R_{\rm e}$ is described by the scatter $\sigma_{ln(R_{\rm e})}$ and the median $<R_{\rm e}>$. More massive galaxies have larger $<R_{\rm e}>$.  The $AR$ distribution of late-type galaxies comes from \cite{padilla08}  and that  of early-type galaxies from \cite{hao06} is adopted.
At a given $AR$ distribution, the other two parameters $n$ and $R_{\rm e}$ vary across the corresponding parameter space. When $n$ falls into $1\le n \le 2.5$ and $2.5 < n \le 6$ , the $AR$ distribution of the late-type and of the early-type galaxies are used, respectively. And $R_{\rm e}$ ranges from 0$\farcs$05 to 0$\farcs$75 (1 to 15 pixels), corresponding to a physical scale of 0.5 to 6.0\,kpc at $z=1$. Similarly, we generate each set of model images for stacking at a fixed $n$ and $AR=1.0$ with $R_{\rm e}$ following the given distribution.
Thirdly, we let $AR$ and $R_{\rm e}$ follow the corresponding distributions and generate models as functions of $n$ and  $<R_{\rm e}>$ to see the integrated effect of the spreads of these parameters.
 Finally, we let $AR$ and $R_{\rm e}$ follow the distributions as described above, and \sersic index $n$ follows a uniform distribution between $1 - 2$ for late-type galaxies and $3 - 4$ for early-type galaxies. This simulates the case that faint galaxies  are usually selected by color (or type) and mass (or luminosity), and their $AR$, $R_{\rm e}$,  and  $n$ often spread over a certain range.
Table~\ref{smp} lists the structural parameters of galaxy models adopted in our simulations.
Figure~\ref{fig:fig1} demonstrates the surface brightness profiles of these single-profile models with $AR=1$ for comparison.

\begin{table}
\begin{center}
\caption{Structural parameters of galaxy models \label{smp}}
\begin{tabular}{@{}cl@{}}
\hline
\hline
\multicolumn{2}{l}{single-profile models} \\
\hline
$n$                 & 1, 1.5, 2.5, 4, 6 \\
$R_{\rm e}$  & 1.0, 2.5, 5.0, 10.0, 15.0\\
$AR$            & 0.10, 0.17, 0.34, 0.50, 0.64, \\
                   & 0.76, 0.86, 0.93, 0.98, 1.00 \\
\hline
\multicolumn{2}{l}{models involving $AR$ spread} \\
\hline
$n$ & 1, 1.5, 2.5, 4, 6\\
$R_{\rm e}$ 	& 1.0, 2.5, 5.0, 10.0, 15.0 \\
\hline
\multicolumn{2}{l}{models involving $R_{\rm e}$ spread} \\
\hline
$n$ & 1, 1.5, 2.5, 4, 6\\
$R_{\rm e,0}$  & 1.0, 2.5,  5.0, 10.0,  15.0\\
$AR$ & 1.0\\
\hline
\multicolumn{2}{l}{models involving $AR$ and $R_{\rm e}$ spreads} \\
\hline
$n$ & 1 , 1.5 , 2.5 , 4 , 6 \\
$R_{\rm e,0}$ & 1.0, 2.5, 5.0, 10.0, 15.0 \\
\hline
\multicolumn{2}{l}{models involving $n$ spread} \\
\hline
$<n>$ & 1.5 , 3.5 \\
$R_{\rm e}$  & 1.0, 2.5, 5.0, 10.0, 15.0 \\
$AR$ & 1.0 \\
\hline
\multicolumn{2}{l}{models involving $AR$, $R_{\rm e}$, and $n$ spreads} \\
\hline
$<n>$ & 1.5 , 3.5 \\
$R_{\rm e,0}$ & 1.0, 2.5, 5.0, 10.0, 15.0 \\
\hline
\hline
\end{tabular}
\end{center}
\tablecomments{(1) The effective radius $R_{\rm e}$ is given in units of pixels (0$\farcs$05 per pixel). Here $R_{\rm e}=15$\,pixels corresponds to a physical scale of 5.4, 6.0 and 6.3\,kpc at $z=[0.7,~1,~2]$, respectively.
(2) The given values of the axis ratio $AR$ correspond approximately to the inclination angle of  90, 80, 70, 60, 50, 40, 30 ,20, 10 and 0 for a disk galaxy. A minimum $AR=0.1$ is chosen when the galaxy is edge-on.
(3) $R_{\rm e,0}$ refers to the median of a log-normal distribution of $R_{\rm e}$.}
\end{table}

\subsection{Stacking galaxy images}

A set of model images for stacking have the same size of  $351\times 351$ pixels, with photon noise and background noise counted.  Before stacking a set of model galaxy images, we first measure the positions of the galaxies using Sextractor \citep{Bertin96} and shift them to the same position in all images. We note that the measurement errors of positions and shifting of images introduce uncertainties into the stacked profile.  We also stack model images free from noise and offsets in position to quantify the corresponding uncertainties due to the noises and errors in aligning the galaxies.
Each set contains  700 galaxy model images and is combined together using the averaging algorithm. Because the position angle of model galaxies is randomly distributed, the averaged profile is rotationally symmetric. The radial surface brightness profile is thus sufficient to characterize the averaged profile. The radial profile is derived from the stacked image using 21 circular apertures with radii from 0.5 to 140\,pixels evenly split in logarithm.  The software tool APER.PRO from the IDL Astronomy User's library \footnote{http://idlastro.gsfc.nasa.gov/} is used to perform aperture photometry.
We use an annulus of $R=6\farcs 24$ to $R=8\farcs 74$ (about 50 to 70\,kpc at $z\sim 1$) for sky background estimation.

\subsection{The Intrinsic Surface Brightness Profile} \label{recover}

A measured surface brightness profile usually needs to correct for the PSF effect in order to obtain the intrinsic profile.
Instead, we follow \citet{Szomoru12} to fit the one-dimensional (1-D) radial surface brightness profile of a stacked image with a library of 1-D \sersic profiles convolved by the same PSF. The library is created in the same way as we generate galaxy models and extract their radial profiles using circular apertures, covering $0\farcs 01 \le R_{\rm e} \le 1\farcs 00$ in a step of $0\farcs 01$ and $0.1\le n \le 8$ in a step of 0.1. The method of least squares is used to select the best fitting and the corresponding \sersic profile is taken as the intrinsic profile for the stacked image.
Similarly, the decomposition of a 1-D surface brightness profile into bulge+disk components can be done by best fitting the profile with the combination of two distinct \sersic profiles.

\section{ANALYSIS AND RESULTS} \label{sec:results}

\subsection{The elongation/inclination effects}\label{ressinlge}

The elongation (or inclination for disk galaxies) of individual stacked galaxies is usually not corrected in stacking. Here we examine how the elongation/inclination affects the recovery of the structural parameters of a \sersic profile. We extract the radial surface brightness profiles using circular apertures from the single-profile galaxy model images listed in  Table~\ref{smp}.   We note that these model images are convolved with ACS PSF.
Figure~\ref{fig:fig2} shows the  single-\sersic profiles as functions of $n$, $R_{\rm e}$ and $AR$. The PSF profile is also presented with an arbitrary normalization for clarity.  Red and blue lines show $AR$=0.34 and 0.68 in each panel. We can see that the surface brightness profile of a galaxy appears to be more compact  at edge-on than at face-on when circular apertures are adopted.  The bias becomes larger for late-type galaxies with larger $R_{\rm e}$.

\begin{figure}
\begin{center}
\includegraphics[width=0.47\textwidth]{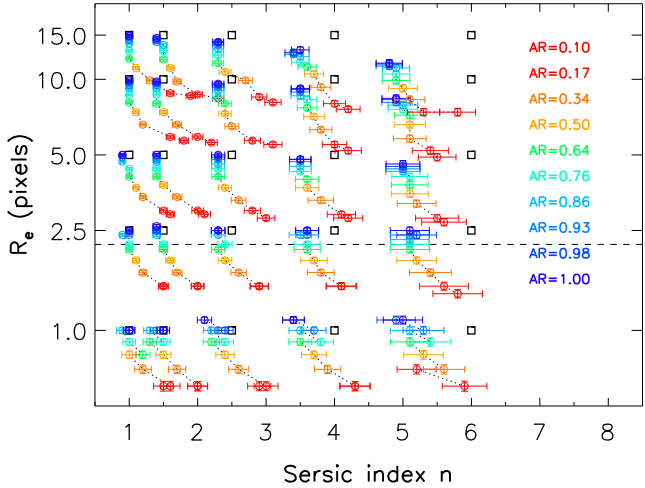}
\caption{Recovery of structural parameters \sersic index $n$ and effective radius $R_{\rm e}$ as a function of axis ratio $AR$. Black squares (barely seen behind blue circles for late-type galaxies) mark the starting points at given ($n$, $R_{\rm e}$). Color-coded circles represent the recovered ($n$, $R_{\rm e}$) at corresponding $AR$ from the starting points (black squares).}
\label{fig:fig3}
\end{center}
\end{figure}

\begin{figure}
\begin{center}
\includegraphics[width=0.47\textwidth]{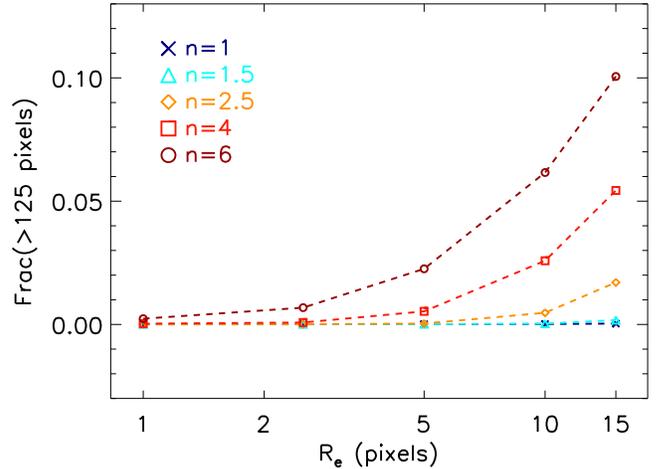}
\caption{The fraction of light out of $R=125$\,pixels as a function of \sersic index $n$ and effective radius $R_{\rm e}$. We can see that the light in the outer regions is marginal for lat-type galaxies, but increases as the $R_{\rm e}$ increases for early-type galaxies. }
\label{fig:fig4}
\end{center}
\end{figure}

We measure the structural parameters from a radial surface brightness profile using the method described in Section~\ref{recover}.  Figure~\ref{fig:fig3} shows the effects of the elongation/inclination on the recovery of the intrinsic structural parameters of these galaxy models. When $AR$=1, the recovered  \sersic index $n$ and effective radius $R_{\rm e}$  perfectly match the input values  when the input \sersic index is low,  but the recovered \sersic index is lower than the input value when the input \sersic index is high. For model galaxies with high \sersic index and large effective radius, the effective radius of the stacked profile tends to be underestimated.
This is because such galaxies exhibit profiles with extended wings out to large radii, leading to an oversubtraction of the background and thus underestimate in both \sersic index and effective radius. We plot the fraction of light outside  $R=$125\,pixels (about 6$\farcs$25 and $\sim$ 50\,kpc at $z\sim$1) as a function of \sersic index $n$ and effective radius $R_{\rm e}$ in Figure~\ref{fig:fig4}. We can see that for late-type galaxies, the light beyond $R=6\farcs 25$ is negligible.  For early-type galaxies, however, the light out of  $R=6\farcs 25$ dramatically increases with the effective radius, and  reaches up to 10\% for galaxies with $n=6$ and $R_{\rm e}$=15\,pixels. This indeed leads to the oversubtraction of background in our measurements, and consequently to underestimate of both \sersic index $n$ and effective radius $R_{\rm e}$ for early-type galaxies.
We note that the underestimate of \sersic index due to background oversubtraction can be corrected once a larger annulus is adopted for background estimate. For a model galaxy with $n=6$ and $R_{\rm e}$=15\,pixels, background estimate from out regions $R>465$\,pixels (186\,kpc at $z=1$) may suppress light contamination from the galaxy to $<$1\%; and the \sersic index can be recovered as well as for galaxies with $n=1$.

 Figure~\ref{fig:fig3} shows that \sersic index $n$ is increasingly overestimated for galaxies with lower $AR$ (i.e., more inclined). The overestimate is significant only at $AR<0.5$, and becomes larger for larger $R_{\rm e}$  at small $n$($n<2.5$),  from $\Delta n \leq$1 when $R_{\rm e}$=15\,pixels (6\,kpc at $z=1$), to $\Delta n \leq$0.5 when $R_{\rm e}$=1\,pixel.  Meanwhile, $R_{\rm e}$ is increasingly underestimated at  decreasing $AR$. The  recovered $R_{\rm e}$ deviates from the intrinsic value by up to 50\% at the minimal $AR$; and the deviation does not strongly depend on $R_{\rm e}$ and $n$.

We conclude that elongation/inclination ($AR$) influences the estimate of structural parameters $n$ and $R_{\rm e}$ when the radial surface brightness profile is extracted from a galaxy image using circular apertures.  The galaxy would appear to be more compact, i.e., with larger $n$ and smaller  $R_{\rm e}$, at decreasing  $AR$,  if the elongation/inclination is ignored.
We point out that the bias in recovering structural parameters through stacking  from elongation effects  is  negligible  for early-type galaxies ($n > 2.5$) because the early-type galaxies usually have $AR>0.5$ \citep[at least  in the local universe,][]{hao06}, and the main source of bias is from background estimation. For late-type galaxies ($n \le 2.5$), the inclination  leads to overestimate of $n$ and underestimate of $R_{\rm e}$.

Next step is to test the measurement of the mean surface brightness profile from the stacked image of individual galaxy models with two parameters fixed and the third parameter following a given distribution.

\subsection{Effects of $AR$ spread}\label{sec:ar}

For each pair of structural parameters ($n$, $R_{\rm e}$) listed in Table~\ref{smp}, a set of 700 galaxy model images are generated to have the fixed $R_{\rm e}$ and $n$ but $AR$ spreading within a distribution. We adopt the $AR$ distribution of late-type galaxies ($n \le 2.5$) from \cite{padilla08} and that of early-type galaxies ($n > 2.5$) from  \cite{hao06}.
These images are stacked together and a radial surface brightness profile is then obtained. The intrinsic parameter of the mean profile is  given by ($n$, $R_{\rm e}$).

\begin{figure*}
\begin{center}
\includegraphics[width=0.82\textwidth]{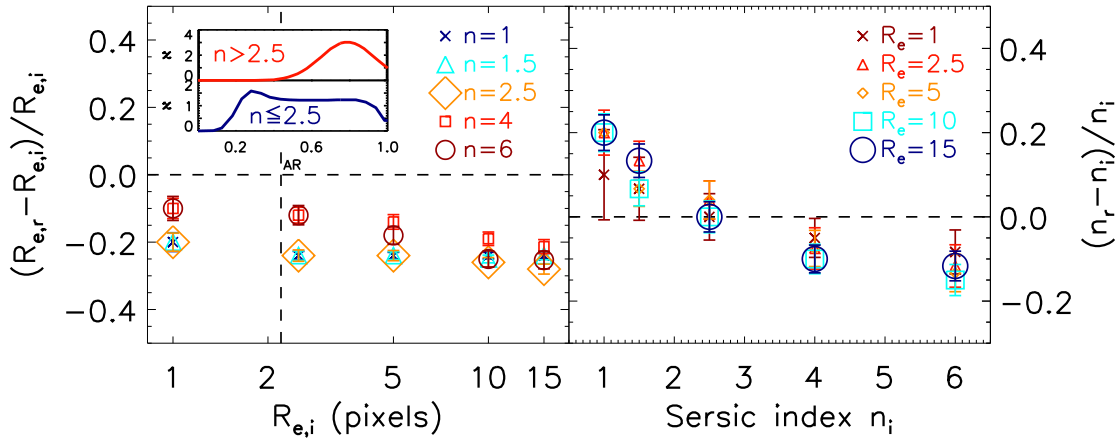}
\caption{Recovery of effective radius $R_{\rm e}$ and \sersic index $n$ through stacking in which the stacked model galaxies spread in $AR$. Left panel shows the deviation of the recovered effective radius $R_{\rm e,r}$ relative to the input effective radius $R_{\rm e,i}$ as functions of $R_{\rm e,i}$ and $n$.  Right panel shows the deviation of the recovered \sersic index $n_{\rm r}$ relative to the input \sersic index $n_{\rm i}$  as functions of $n_{\rm i}$ and $R_{\rm e}$. The inner panels show the adopted $AR$ distribution of early-type galaxies ($n>2.5$) from \cite{hao06} and of late-type galaxies ($n\le 2.5$) from \cite{padilla08}. }
\label{fig:fig5}
\end{center}
\end{figure*}

\begin{figure*}
\begin{center}
\includegraphics[width=0.82\textwidth]{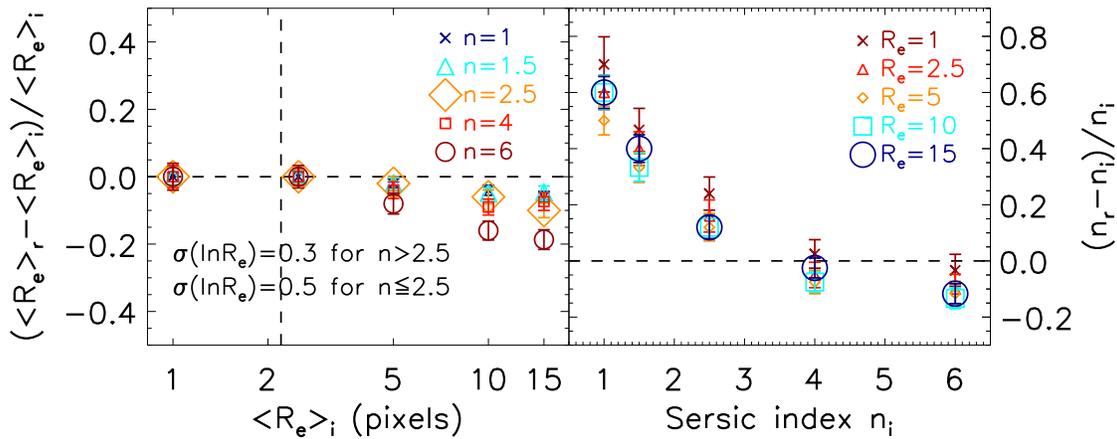}
\caption{Recovery of median effective radius $<R_{\rm e}>$ and \sersic index $n$ by stacking model galaxies with $R_{\rm e}$ satisfying the log-normal distribution from \cite{shen03}.  Left plot shows the deviation of the recovered effective radius $<R_{\rm e}>_{\rm r}$ relative to the input median $R_{\rm e,0}$ as functions of $R_{\rm e,0}$ and input \sersic index $n_{\rm i}$.   Right plot shows the deviation of the recovered \sersic index $n_{\rm r}$ relative to $n_{\rm i}$  as functions of $n_{\rm i}$ and $R_{\rm e,0}$. Here $AR$ is fixed to unity (face-on). } 
\label{fig:fig6}
\end{center}
\end{figure*}

 Figure~\ref{fig:fig5} shows the results of our simulation when  $AR$ is spread according to a realistic distribution rather than fixed. The two empirical $AR$ distributions are shown in the inner panels. The left plot gives the difference between recovered $R_{e}$ and input $R_{\rm e}$ as a function of the input $R_{\rm e}$ and the right plot presents the deviation of recovered $n$ from input $n$ as a function of the input $n$.
The recovered $R_{\rm e}$ is systematically smaller by 12\% to 27\% over $1\le R_{\rm e} \le 15$\,pixels.
For late-type galaxies, the degree of underestimate in $R_{\rm e}$ does not depend on $R_{\rm e}$ itself, suggesting that the $AR$ distribution is responsible for the underestimate.
But for early-type galaxies, the degree of underestimate in $R_{\rm e}$ increases as the $R_{\rm e}$ increases.
 Given that the oversubtraction of background becomes more serious for early-type galaxies with larger size,  it is reasonable to attribute the increase of the degree of underestimate in  $R_{\rm e}$ to the oversubtraction of background.
Apparently, the estimate of $R_{\rm e}$ is more biased for late-type galaxies due to the highly-inclined ones ($AR<0.5$) which are absent in the early-type galaxies when the effective radius is not very large.
Similarly,  $n$ is overestimated by up to 20\%  for late-type galaxies because of the inclination effect due to those with $AR<0.5$. Again, the overestimate of $n$ does not rely on $R_{\rm e}$  significantly. For early-type galaxies,  $n$ can be underestimated up to 15\%, and the underestimation of $n$ increases as $n$ increases. As most of them are with $AR>$0.5 in terms of the $AR$ distribution from \cite{hao06}, the overestimation of $n$ can be ignored for early-type galaxies, thus the oversubtraction of background dominates the estimation of $n$ and makes $n$ underestimated, especially for galaxies with larger $n$ which have more extended halo in the outskirts.
As shown in Figure~\ref{fig:fig3}, the recovery of $n$ is marginally biased by inclination/elongation effects at $AR>$0.5.  These results denote that the function of the $AR$ distribution regulates the deviation of the recovered $R_{\rm e}$ and $n$ from the original values for late-type galaxies, and the estimation of background affects the estimation of structural parameters for early-type galaxies.

Differing from the claim in \cite{vdk10} that the recovery of structural parameters is not sensitive to the distribution of $AR$, we show that the $AR$ distribution may significant bias the recovery of $R_{\rm e}$ and $n$, dependent on the fraction of highly-inclined or elongated ones;  the stacked profile appears to be more compact (i.e., with smaller $R_{\rm e}$ and  higher $n$ at the same time) if the elongation/inclination effect is not corrected.

\begin{figure*}
\begin{center}
\includegraphics[width=0.82\textwidth]{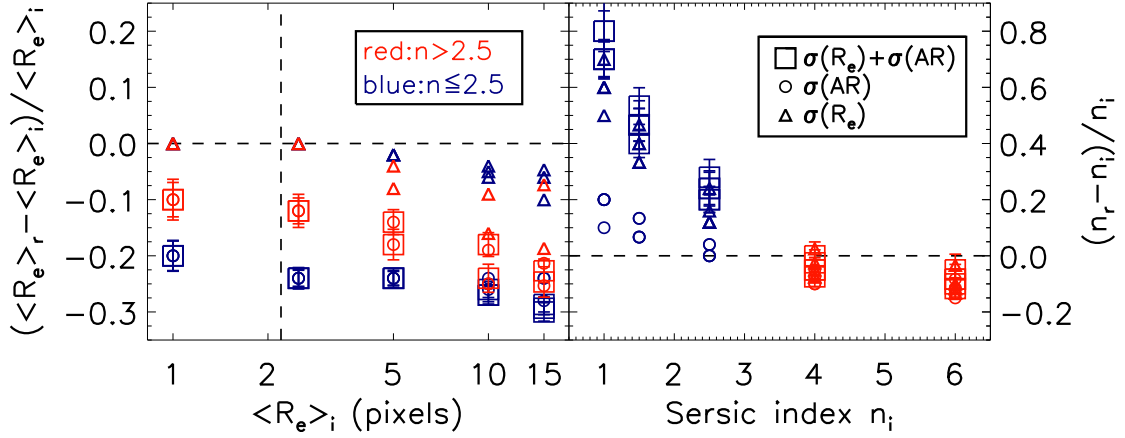}
\caption{Recovery of structural parameters $R_{\rm e}$ and $n$ through stacking model galaxies with $AR$ and $R_{\rm e}$ spreads counted. Left plot shows the deviation of recovered mean $<R_{\rm e}>_{\rm r}$ relative to the input median $R_{\rm e,0}$ as a function of $R_{\rm e,0}$. Right plot shows the deviation of recovered \sersic index $n_{\rm r}$ relative to the input $n_{\rm i}$ as a function of $n_{\rm i}$.  Squares represent the results when the two spreads are involved.  Circles mark the results from Figure~\ref{fig:fig5} that only the $AR$ spread is counted. Triangles show the results when the $R_{\rm e}$ spread is included (from Figure~\ref{fig:fig6}). Blue and red colors correspond to $n\le2.5$ and $n>2.5$, respectively.}
\label{fig:fig7}
\end{center}
\end{figure*}

\subsection{Effects of $R_{\rm e}$ spread}\label{sec:re}

Figure~\ref{fig:fig6} shows  our simulation results of stacking model galaxies with a fixed $n$ but $R_{\rm e}$ following a log-normal distribution. Here $AR=1$ is adopted to get rid of the $AR$ effect.   The log-normal distribution is  described by the median $R_{\rm e,0}$ and scatter $\sigma_{\ln(R_{\rm e})}$. We adopt $\sigma_{\ln(R_{\rm e})}$=0.3\,dex for early-type galaxies and $\sigma_{\ln(R_{\rm e})}$=0.5\,dex for late-type galaxies from \cite{shen03}.
As shown in the left panel of Figure~\ref{fig:fig6},  the input median effective radius $R_{\rm e,0}$ is well recovered through stacking  for late-type galaxies and early-type galaxies with small effective radius, but for early-type galaxies with large radius, the underestimation of $R_{\rm e}$ can be up to 15\% due to the oversubtraction of background. However,   $n$ is increasingly overestimated by up to 60\% at decreasing $n$. This tendency has no dependence on the median of the log-normal distribution of $R_{\rm e}$. For early-type galaxies, the $n$ can still be underestimated due to the oversubtraction of background.

One can  infer from Figure~\ref{fig:fig6} that the median of a log-normal distribution of $R_{\rm e}$ can be properly measured from the stacked profile of galaxies; but the spread of $R_{\rm e}$ leads to an overestimate of $n$ for late-type galaxies ($n\leq$2.5).

\begin{figure*}
\begin{center}
\includegraphics[scale=0.78]{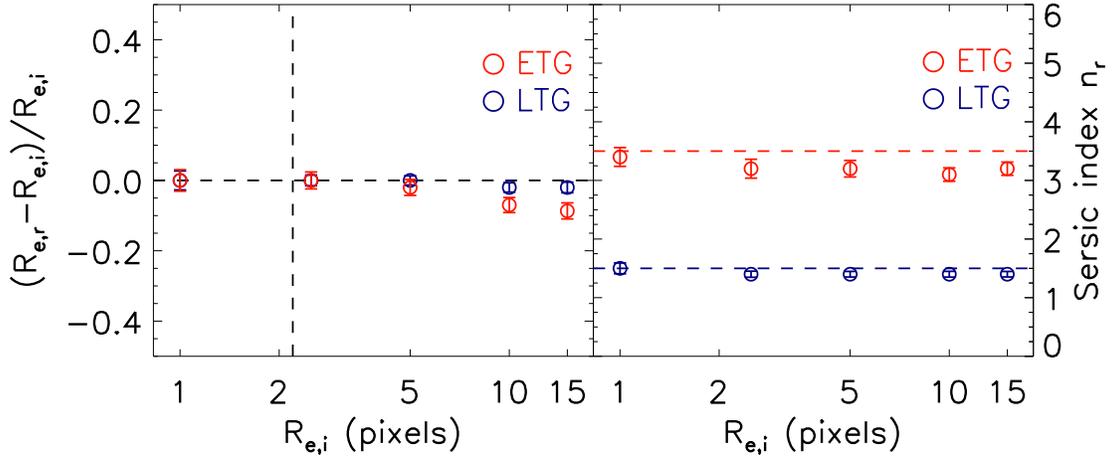}
\caption{Recovery of structural parameters $R_{\rm e}$ and $n$ through stacking galaxies with \sersic index $n$ uniformly distributed between $1-2$ for late-type galaxies and $3-4$ for early-type galaxies. The effective radius is fixed at given value in each stacking set and the axis ratio $AR=1$ is chosen for all sets. The left panel shows the deviation of the recovered effective radius $R_{\rm e,r}$ relative to the input value $R_{\rm e,i}$ as a function of $R_{\rm e,i}$ for the two types of galaxies. The right panel shows the deviation of the recovered \sersic index $n_{\rm r}$ relative to the median input $n_{\rm i}$ as a function of the effective radius $R_{\rm e,i}$.}
\label{fig:fig8}
\end{center}
\end{figure*}

\begin{figure*}
\begin{center}
\includegraphics[scale=0.78]{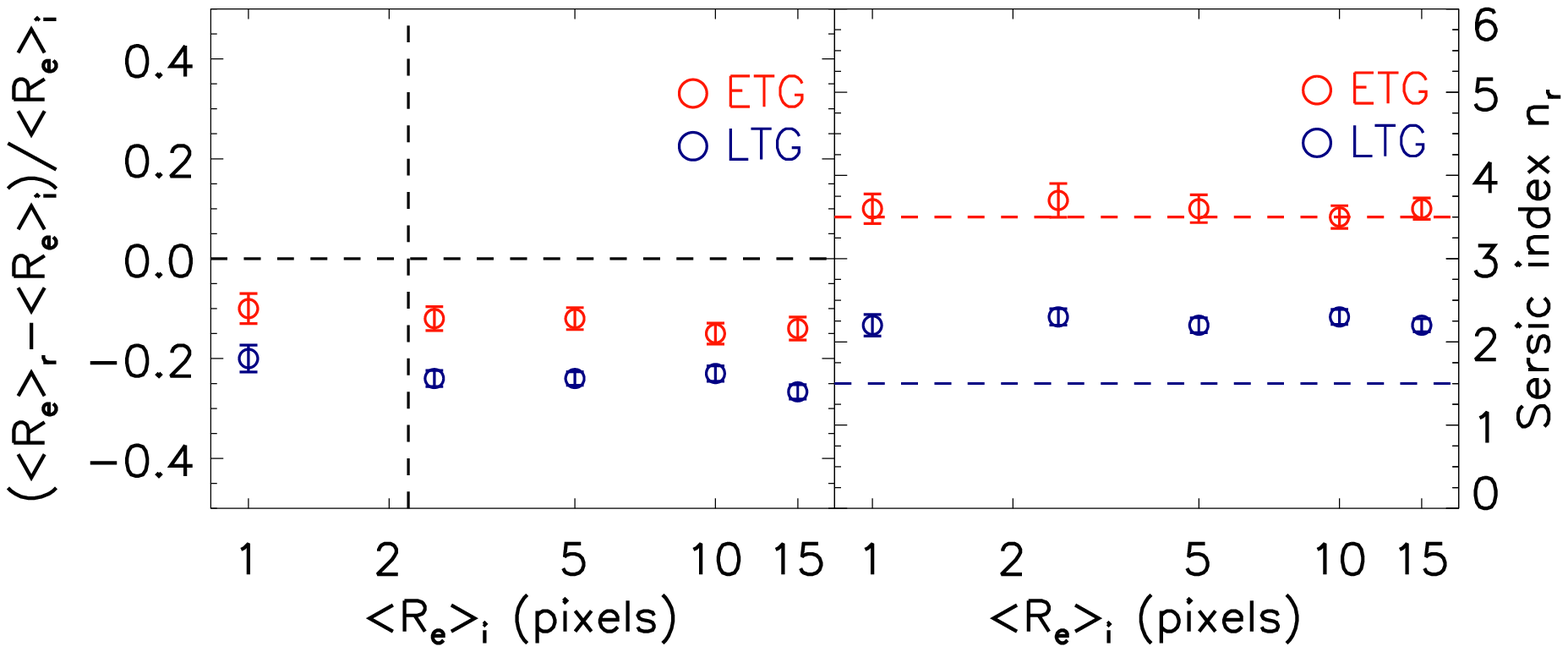}
\caption{Recovery of structural parameters $R_{\rm e}$ and $n$ through stacking galaxies with spreads of all  parameters ($AR$, $R_{\rm e}$,$n$) counted. The left panel shows the deviation of recovered effective radius $<R_{\rm e}>_{\rm r}$ relative to the input median effective radius $<R_{\rm e}>_{\rm i}$ as a function of $<R_{\rm e}>_{\rm i}$ for different type of galaxies. The right panel shows the deviation of recovered \sersic index $n$ relative to the median input median $n$ as a function of the effective radius for different type of galaxies.}
\label{fig:fig9}
\end{center}
\end{figure*}

\subsection{Effects of $AR$ and $R_{\rm e}$ spreads} \label{sec:are}

We have shown that the spread of $AR$ biases the estimates of structural parameters of the mean profile through stacking, leading the stacked profile to be more compact, say with smaller $R_{\rm e}$ and higher $n$; and the spread of $R_{\rm e}$ does not influence the estimate of median $R_{\rm e}$ but  deviates $n$ to be higher. The effects of the two spreads are  significant only for stacking of late-type galaxies ($n\leq$2.5).
Now we include both the two spreads in stacking and examine their effects on the recovery of the structural parameters.
Again, $AR$ and $R_{\rm e}$ follow distinct distributions for early- and late-type galaxies as mentioned before. We note that the effects of the two spreads are not correlated with the median $R_{\rm e,0}$. 

Figure~\ref{fig:fig7} shows the results of stacking with both of the $AR$ and $R_{\rm e}$ spreads involved.  We also over plot the results from Figure~\ref{fig:fig5} and \ref{fig:fig6} for comparison.
We can see that the mixture of $AR$ and $R_{\rm e}$ spreads biases the estimates of median $R_{\rm e,0}$ and $n$ in the same way as the effects of the two spreads combine linearly together.  The median $R_{\rm e,0}$ is underestimated by  20\%-27\% and 10\%-20\% for late-type and early-type galaxies, depending on the $R_{\rm e}$ of the galaxies, respectively, caused by the $AR$ spread and the estimation of background; and $n$ is increasingly overestimated by up to 70\% at decreasing $n$, equal to a linear combination of the deviations caused by each of the two spreads.

In summary, our simulations manifest that the measured size ($R_{\rm e}$) and \sersic index ($n$) of the averaged profile by stacking a set of galaxies deviate from the input values when the stacked galaxies disperse in axis ratio ($AR$) and/or half-light radius ($R_{\rm e}$). The deviations depend on the type of stacked galaxies ($n$) and distribution functions of $AR$ and $R_{\rm e}$. The stacked profile tends to be more compact for late-type galaxies,
and the oversubtraction of background dominates the estimation of structural parameters for early-type galaxies and make the profile to be smaller in both $R_{\rm e}$ and $n$.
With given distributions of  $AR$ and/or $R_{\rm e}$, the deviations in estimates of $R_{\rm e}$ and $n$ can be quantitatively determined and thus corrected accordingly.

It is worth noting that the effects caused by the spread of $n$ depend strongly on the distribution function of $n$. The averaged profile of a set of \sersic profiles with fixed $R_{\rm e}$ and $AR$=1 is much close to the median one. The derived $n$ from stacking are rather reliable with uncertainties of $<$0.5 for both late-type and early-type galaxies \citep[see the Appendix of][for more details]{vdk10}. We will also discuss this issue in Section~\ref{sec:n}.
 However, it remains to be explored when two distinct types of profiles are stacked together.

\subsection{Effects of $n$ spread} \label{sec:n}

Generally speaking, a set of galaxies of similar properties (e.g., stellar mass or color) are often selected for stacking to derive their mean profile. For instance, galaxies are often divided into two populations: star-forming and quiescent. The star-forming galaxies tend to have Sersic index between 1 and 2 and the quiescent galaxies have Sersic index systematically higher.   In order to test how the spread of $n$ affects the stacked results, we adopt a uniform distribution of $n$ between $1-2$ for late-type galaxies and $3-4$ for early-type galaxies, with fixed $R_{\rm e}$ and $AR=1.0$.   Figure~\ref{fig:fig8} shows that the $R_{\rm e}$ is almost identical to the input values for LTGs and small ETGs, and for large ETGs, the $R_{\rm e}$ can be slightly underestimated due to the oversubtraction of background, which can be seen from Figure~\ref{fig:fig3}. The median \sersic index can be recovered well for both late-type galaxies and early-type galaxies, though the recovered \sersic index is slightly lower than the input median \sersic index for the early-type galaxies with large effective radius.

\subsection{Effects of spreads in $AR$, $R_{\rm e}$ and $n$} \label{sec:all}

We account for the spreads in  $AR$, $R_{\rm e}$ and $n$ together in our stacking exercises, and examine whether the averaged structural parameters can be well recovered through stacking. Our results shown in Figure~\ref{fig:fig9} denote that the $R_{\rm e}$ may be underestimated up to 20\%$-$26\% for late-type galaxies but only 10\%$-$15\% for early-type galaxies.   The presence of highly-inclined late-type galaxies is believed to cause the additional bias to the underestimate of the averaged effective radius. The median $n$ can be well recovered for early-type galaxies, but overestimated by $\delta n \sim 0.7$ for late-type galaxies. This is due to late-type galaxies having larger dispersion in $R_{\rm e}$ and highly-inclined ones. The early-type galaxies have smaller dispersion in $R_{\rm e}$, leading to a smaller overestimate of $n$, which can be diluted by the underestimation of $n$ caused by the oversubtraction of background.

\begin{figure*}
\begin{center}
\includegraphics[scale=0.75]{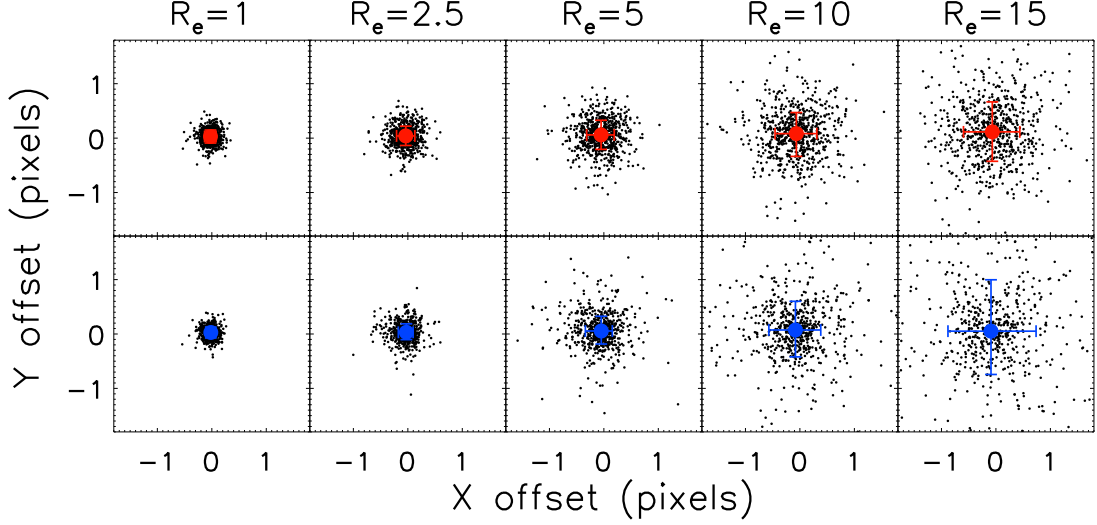}
\caption{Uncertainties of the measured position for model galaxies with $i = 24-24.75$\,mag as a function of effective radius.  The errorbars indicate the 16th and 84th percentiles of the distribution in each stacking set.}
\label{fig:fig10}
\end{center}
\end{figure*}

\begin{figure*}
\begin{center}
\includegraphics[scale=0.78]{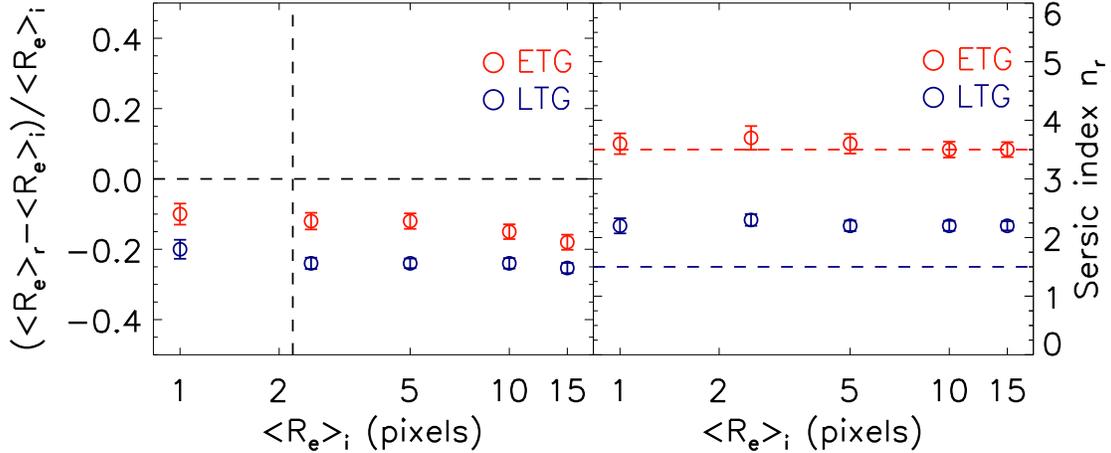}
\caption{Recovery of structural parameters $R_{\rm e}$ and $n$ through stacking galaxies with spreads of all parameter ($AR$,$R_{\rm e}$,$n$) counted and model galaxies exactly aligned to the same position in all images. The left panel shows the deviation of recovered effective radius $<R_{\rm e}>_{\rm r}$ relative to the input median effective radius $<R_{\rm e}>_{\rm i}$ as a function of $<R_{\rm e}>_{\rm i}$ for different type of galaxies. The right panel shows the deviation of recovered \sersic index $n$ relative to the median input median $n$ as a function of the effective radius for different type of galaxies. We can see that There is no significant difference compared with Figure~\ref{fig:fig9}, indicating that the centering offset influences little on the recovered structural parameters.}
\label{fig:fig11}
\end{center}
\end{figure*}

\subsection{Uncertainties in aligning images} \label{sec:offset}

 For faint galaxies, the measured position is affected by noise and thus offset slightly from the true position.  In our stacking analysis, we measure the positions of model galaxies using Sextractor and shift the galaxies to the same position in all images before stacking. Figure~\ref{fig:fig10} shows that 68 percentile of the measured positions deviate less than  1\,pixel from the true positions.  At fixed magnitudes, galaxies with more extended structure (i.e., larger effective radius) exhibit larger deviation in measuring their positions.   In order to quantitatively estimate the uncertainties caused by the errors in aligning images, we repeat the stacking exercises presented in Figure~\ref{fig:fig9} but with model galaxies generated to exactly center at the same position in all images. No image aligning is employed here. Figure~\ref{fig:fig11} shows the corresponding results. We can see that the recovered parameters of the stacked profiles at all $R_{\rm e}$ and $n$ from Figure~\ref{fig:fig11} are nearly identical to these given in Figure~\ref{fig:fig9}, indicating that the error in aligning galaxy images is not a source to bias the recovery of the mean structural parameters through stacking, at least for galaxies with $24-24.75$\,mag and brighter.

\begin{table*}
\begin{center}
\centering \scriptsize
\caption{Recovery of structural parameters through stacking bulge+disk galaxy models \label{tab:2com}}
\begin{tabular}{ccccccccccccccccccccc}
\hline
\hline
\multicolumn{3}{c}{Input parameters} && \multicolumn{17}{c}{Recovered parameters} \\
\cline{5-21} \\
\multicolumn{3}{c}{} && \multicolumn{5}{c}{Individual bulge+disk models} &&  \multicolumn{5}{c}{Stacks with $f(AR_{\rm D})$ \& $\sigma(\ln\,R_{\rm e,D})$} && \multicolumn{5}{c}{Stacks with $f(AR_{\rm D})$,  $\sigma(\ln\,R_{\rm e,D})$ \& $\sigma(\ln\,R_{\rm e,B})$}\\
\cline{1-3}\cline{5-9}\cline{11-15}\cline{17-21} \\
$R_{\rm e,D}$ & $R_{\rm e,B}$ & $B/T$ & & $R_{\rm e,D}$ & $R_{\rm e,B}$ & $B/T$  & $n_{\rm T}$ & $R_{\rm e,T}$ &  & $R_{\rm e,D}$ & $R_{\rm e,B}$ & $B/T$  & $n_{\rm T}$ & $R_{\rm e,T}$ &  & $R_{\rm e,D}$ & $R_{\rm e,B}$ & $B/T$  & $n_{\rm T}$ & $R_{\rm e,T}$\\
\cline{1-3}\cline{5-9}\cline{11-15}\cline{17-21} \\
15.0 & 15.0 & 0.1 && 15.0 & 15.5 & 0.10 & 1.2 & 14.6 && 10.4 & 15.9 & 0.50 & 1.9 & 11.4 &&   10.8 & 13.6 & 0.46 & 1.8 & 11.0 \\
15.0 & 10.0 & 0.1 && 15.0 & 10.2 & 0.10 & 1.3 & 14.2 && 10.6 & 13.2 & 0.46 & 1.8 & 10.8 &&   10.3 & 15.1 & 0.52 & 1.9 & 11.1 \\
15.0 &  5.0 & 0.1 && 15.0 &  5.0 & 0.10 & 1.4 & 13.2 && 10.1 & 13.3 & 0.53 & 2.0 & 10.6 &&   10.4 & 12.1 & 0.51 & 1.9 & 10.4 \\
10.0 & 10.0 & 0.1 && 10.0 & 10.1 & 0.10 & 1.1 & 9.7  &&  7.6 &  8.1 & 0.45 & 1.8 &  7.4 &&    7.6 &  8.1 & 0.45 & 1.8 &  7.4 \\
10.0 &  5.0 & 0.1 && 10.0 &  5.0 & 0.10 & 1.2 & 9.2  &&  7.5 &  7.3 & 0.46 & 1.8 &  7.1 &&    7.5 &  7.3 & 0.46 & 1.8 &  7.1 \\
10.0 & 2.5 & 0.1 && 10.0 & 2.5 & 0.10 & 1.4 & 8.7 && 7.4 & 6.6 & 0.49 & 1.9 & 6.8  && 7.5  & 6.4  & 0.49 & 1.9 & 6.8  \\
 5.0 &  5.0 & 0.1 &&  5.0 &  5.0 & 0.10 & 1.2 &  4.9 &&  4.2 &  3.7 & 0.50 & 1.9 &  3.8 &&    4.1 &  3.9 & 0.50 & 1.9 &  3.8 \\
5.0 & 2.5 & 0.1 && 5.0 & 2.5 & 0.10 & 1.2 & 4.7 && 4.1 & 3.4 & 0.51 & 1.9 & 3.7  && 4.1  & 3.4  & 0.51 & 1.9 & 3.7  \\
5.0 & 1.0 & 0.1 && 5.0 & 1.0 & 0.10 & 1.2 & 4.4 && 4.1 & 3.0 & 0.54 & 2.0 & 3.5  && 4.1  & 3.0  & 0.54 & 2.0 & 3.5  \\
\cline{1-3}\cline{5-9}\cline{11-15}\cline{17-21}                                             \\
15.0 & 15.0 & 0.5 && 15.0 & 15.0 & 0.49 & 2.1 & 13.8 && 10.6 & 14.2 & 0.65 & 2.3 & 11.6 &&   10.6 & 13.8 & 0.67 & 2.4 & 11.6 \\
15.0 & 10.0 & 0.5 && 15.0 & 9.9 & 0.49 & 2.4 & 12.1 && 10.3 &  10.8 & 0.67 & 2.4 & 10.1 &&   10.4 & 10.7 & 0.68 & 2.5 & 10.1 \\
15.0 &  5.0 & 0.5 && 15.0 &  5.0 & 0.50 & 2.9 &  9.0 && 10.1 &  7.0 & 0.74 & 2.8 &  7.8 &&   10.2 &  7.1 & 0.76 & 2.8 &  7.8 \\
10.0 & 10.0 & 0.5 && 10.0 & 10.0 & 0.49 & 2.0 &  9.3 &&  7.5 &  9.3 & 0.67 & 2.3 &  7.9 &&    7.1 &  9.8 & 0.70 & 2.5 &  8.0 \\
10.0 &  5.0 & 0.5 && 10.0 &  5.0 & 0.50 & 2.2 &  7.0 &&  7.2 &  5.8 & 0.68 & 2.5 &  6.2 &&    7.1 &  5.9 & 0.70 & 2.5 &  6.2 \\
10.0 & 2.5 & 0.5 && 10.0 & 2.5 & 0.50 & 2.8 & 5.3 && 7.2 & 3.9 & 0.77 & 2.9 & 4.7  && 7.3  & 3.9  & 0.77 & 2.9 & 4.7  \\
 5.0 & 5.0 & 0.5 &&  5.0 &  5.0 & 0.50 & 2.1 & 4.9 &&  3.9 &  4.8 & 0.70 & 2.6 &  4.1 &&    3.7 &  5.1 & 0.70 & 2.7 &  4.1 \\
5.0 & 2.5 & 0.5 && 5.0 & 2.5 & 0.50 & 2.1 & 3.6 && 3.9  & 2.9 & 0.70 & 2.5 & 3.2  && 4.0  & 2.9  & 0.72 & 2.6 & 3.2  \\
5.0 & 1.0 & 0.5 && 5.0 & 1.0 & 0.50 & 2.2 & 2.6 && 4.2  & 1.6 & 0.76 & 2.9 & 2.2  && 4.0  & 1.8  & 0.80 & 2.9 & 2.3  \\
\cline{1-3}\cline{5-9}\cline{11-15}\cline{17-21}                                             \\
15.0 & 15.0 & 0.9 && 14.7 & 15.1 & 0.90 & 3.5 & 14.5 && 10.9 & 13.8 & 0.87 & 3.2 & 12.7 &&   9.3 & 14.2 & 0.91 & 3.3 & 12.7 \\
15.0 & 10.0 & 0.9 && 15.0 & 10.0 & 0.90 & 3.6 & 10.3 && 9.2 & 9.8 & 0.89 & 3.2 & 9.3 &&   7.1 & 10.3 & 0.92 & 3.4 & 9.4 \\
15.0 &  5.0 & 0.9 && 15.0 &  5.0 & 0.90 & 3.9 &  5.7 && 7.6 &  5.1 & 0.91 & 3.4 &  5.3 &&   4.8 &  5.6 & 0.96 & 3.7 &  5.5 \\
10.0 & 10.0 & 0.9 && 10.0 & 10.0 & 0.90 & 3.5 &  9.8 &&  6.4 & 10.2 & 0.89 & 3.3 &  9.0 &&    5.3 &  10.5 & 0.90 & 3.5 &  9.1 \\
10.0 &  5.0 & 0.9 && 10.0 &  5.0 & 0.90 & 3.6 &  5.4 &&  6.3 &  5.0 & 0.89 & 3.3 &  5.1 &&    4.3 &  5.5 & 0.93 & 3.6 &  5.2 \\
10.0 & 2.5 & 0.9 && 10.0 & 2.5 & 0.90 & 3.8 & 3.0 && 5.2  & 2.7 & 0.93 & 3.6 & 2.9  && 6.4  & 2.7  & 0.95 & 3.7 & 2.9  \\
 5.0 &  5.0 & 0.9 &&  5.0 &  5.0 & 0.90 & 3.6 &  4.9 &&  4.6 &  4.8 & 0.88 & 3.3 &  4.6 &&    3.8 &  5.0 & 0.90 & 3.5 &  4.7 \\
5.0 & 2.5 & 0.9 && 5.0 & 2.5 & 0.90 & 3.5 & 2.7 && 3.9  & 2.5 & 0.89 & 3.3 & 2.7  && 4.1  & 2.5  & 0.91 & 3.4 & 2.7  \\
5.0 & 1.0 & 0.9 && 5.0 & 1.0 & 0.90 & 3.5 & 1.3 && 3.4  & 1.1 & 0.94 & 3.7 & 1.2  && 4.0  & 1.0  & 0.94 & 3.6 & 1.3  \\
\hline
\hline
\end{tabular}
\end{center}
\end{table*}

\begin{figure*}
\begin{center}
\includegraphics[width=0.75\textwidth]{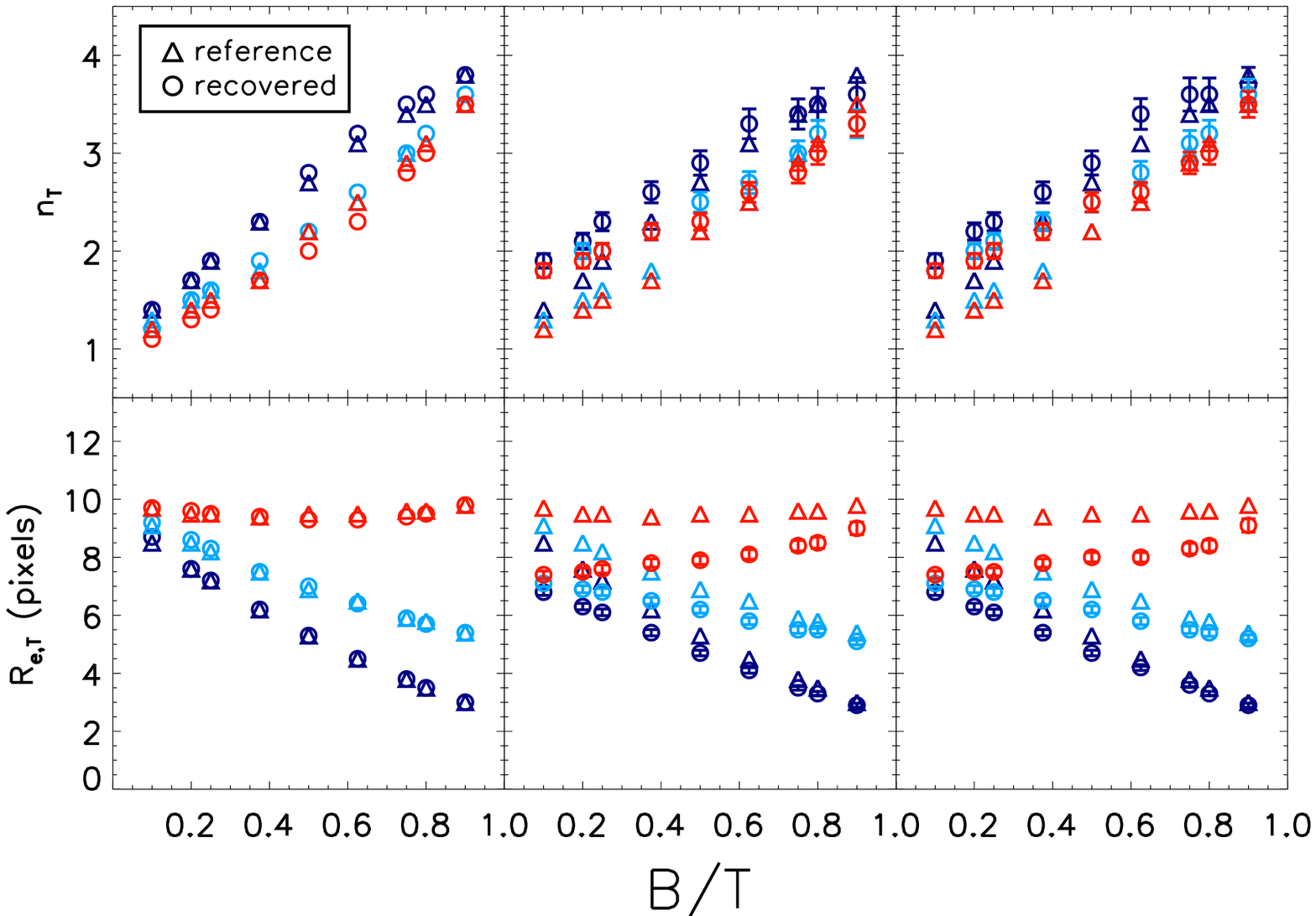}
\caption{Global structural parameters  $n_{\rm T}$ and $R_{\rm e,T}$  as functions of $B/T$ and the ratio of $R_{\rm e,B}$ to $R_{\rm e,D}$ for individual bulge+disk models ({\it left}),  stacked profiles of bulge+disk models with spreads in $AR_{\rm D}$ and $R_{\rm e,D}$ ({\it middle}) counted, and  stacked profiles of bulge+disk models with spreads in $AR_{\rm D}$ and $R_{\rm e,D}$ and $R_{\rm e,B}$ included  ({\it right}). By default, $n=4$ is adopted for the bulge and $n=1$ for the disk.  Color codes $R_{\rm e,B}/R_{\rm e,D}$= 0.25 (blue), 0.5 (cyan) and 1 (red) with $R_{\rm e,D}$=10\,pixels.  Circles represent the recovered profiles from PSF-convolved models, while triangles in all panels mark the reference profiles from the individual PSF-free models.}
\label{fig:fig12}
\end{center}
\end{figure*}

\subsection{Stacking  bulge+disk galaxies}\label{subsec:two}

The composite-type galaxies are commonly composed of two distinct \sersic components, i.e., bulge+disk.  We generate bulge+disk galaxy models to test the recovery of structural parameters through stacking such objects.  We assume that bulges are  classical (i.e., de Vaucouleurs  type with $n=4$) and disks are exponential ($n=1$), as often used in morphological studies of high-$z$ galaxies \citep[e.g.,][]{Bruce12,Lang14}.  The effective radius of the bulges $R_{\rm e,B}$ is set not to be larger than the effective radius of the disks $R_{\rm e,D}$.  Here we adopt  $R_{\rm e,D}$=[5,10,15]\,pixels and $R_{\rm e,B}$ is chosen accordingly, as listed in Table~\ref{tab:2com}. A new parameter bugle-to-total light ratio ($B/T$) is used to measure the fraction between the two components in a galaxy.  The $B/T$ is set to be 0.1, 0.2, 0.25, 0.375, 0.5, 0.625, 0.75, 0.8 and 0.9.

We firstly fit the 1-D profiles of individual noise-free bulge+disk models with single \sersic profiles, selecting the recovered profiles of best-fit parameters $n_{\rm T}$ and $R_{\rm e,T}$.   Meanwhile, we also obtain the corresponding single \sersic profiles best fitting the 1-D profiles of the bulge+disk models without PSF convolution using the method presented in Section~\ref{recover}. The best-fit profiles to the PSF-free models are taken as the reference profiles. Here $AR$=1 is adopted for all galaxy models.  We do not include noises in these bulge+disk decomposition exercises in order to test how well the actual parameters of bulge+disk models can be derived from the 1-D profile fitting.
Secondly, we examine the estimate of structural parameters through stacking bulge+disk models with spreads in $AR$ and $R_{\rm e}$ counted only for the disk component, aiming at addressing how the recovered structural parameters depend on the input model parameters.
We stack images of bulge+disk models of given ($R_{\rm e,B}$, $R_{\rm e,D}$, $B/T$) with spreads in $AR_{\rm D}$ and $R_{\rm e,D}$.  The $AR_{\rm D}$ spread in Section~\ref{sec:ar} and the spread of $R_{\rm e}$ for late-type galaxies in Section~\ref{sec:re} are adopted.
Thirdly, the spread of $R_{\rm e,B}$ for the bulge component and the $AR_{\rm D}$ and $R_{\rm e,D}$ spreads for the disk component are taken into account in stacking bulge+disk models of given ($R_{\rm e,B}$, $R_{\rm e,D}$, $B/T$). Similarly, the best-fit \sersic profile to the 1-D profile of the stacked image is taken as the recovered profile and the reference profile is obtained from the best-fit profile with removal of noise and PSF effect. Table~\ref{tab:2com} presents the recovered structural parameters in above three cases for three representative values $B/T=$0.1, 0.5 and 0.9.
These stacking processes deal with the model galaxy images with photon noise and background included to match the actual observations.

\begin{table*}
\begin{center}
\centering \scriptsize
\caption{Recovery of structural parameters through stacking pseudo-bulge ($n=2.5$)+disk galaxy models. \label{tab:pseudo}}
\begin{tabular}{ccccccccccccccccccccc}
\hline
\hline
\multicolumn{3}{c}{Input Parameters} && \multicolumn{17}{c}{Recovered Parameters} \\
\cline{5-21} \\
\multicolumn{3}{c}{} && \multicolumn{5}{c}{Individual bulge+disk models} &&  \multicolumn{5}{c}{Stacks with $f(AR_{\rm D})$ \& $\sigma(\ln\, R{\rm e,D})$} && \multicolumn{5}{c}{Stacks with $f(AR_{\rm D})$,  $\sigma(\ln\,R_{\rm e,D})$ \& $\sigma(\ln\,R_{\rm e,B})$}\\
\cline{1-3}\cline{5-9}\cline{11-15}\cline{17-21} \\
$R_{\rm e,D}$ & $R_{\rm e,B}$ & $B/T$ & & $R_{\rm e,D}$ & $R_{\rm e,B}$ & $B/T$  & $n_{\rm T}$ & $R_{\rm e,T}$ &  & $R_{\rm e,D}$ & $R_{\rm e,B}$ & $B/T$  & $n_{\rm T}$ & $R_{\rm e,T}$ &  & $R_{\rm e,D}$ & $R_{\rm e,B}$ & $B/T$  & $n_{\rm T}$ & $R_{\rm e,T}$\\
\cline{1-3}\cline{5-9}\cline{11-15}\cline{17-21} \\
15.0 & 15.0 & 0.1 && 14.9 & 20.0 & 0.07 & 1.1 & 14.8 && 10.9 & 14.3 & 0.44 & 1.8 & 11.3 && 10.9 & 13.8 & 0.44 & 1.8 & 11.2  \\
15.0 & 10.0 & 0.1 && 14.7 & 14.5 & 0.09 & 1.2 & 14.3 && 10.5 & 14.3 & 0.45 & 1.8 & 11.0 && 10.7 & 13.3 & 0.45 & 1.8 & 10.9  \\
15.0 &  5.0 & 0.1 && 14.6 &  7.9 & 0.12 & 1.4 & 13.3 && 10.0 & 13.1 & 0.50 & 1.9 & 10.4 && 10.1 & 13.0 & 0.51 & 1.9 & 10.4  \\
10.0 & 10.0 & 0.1 && 10.0 & 11.7 & 0.07 & 1.1 & 9.9  && 7.6 &  8.5 & 0.43 & 1.7 &  7.5 &&  7.6 &  8.4 & 0.43 & 1.7 &  7.5  \\
10.0 &  5.0 & 0.1 &&  9.8 &  6.1 & 0.09 & 1.2 & 9.3  && 7.4 &  7.4 & 0.43 & 1.7 &  7.1 &&  7.4 &  7.5 & 0.44 & 1.8 &  7.1  \\
10.0 & 2.5 & 0.1 && 9.8 & 3.6  & 0.12 & 1.3 & 8.6 && 7.3  & 6.5 & 0.48 & 1.9 & 6.8  && 7.3  & 6.6  & 0.49 & 1.9 & 6.8  \\
 5.0 &  5.0 & 0.1 &&  5.1 &  4.1 & 0.07 & 1.1 &  5.0 &&  4.2 &  3.8 & 0.48 & 1.8 &  3.9 &&  4.2 &  3.8 & 0.49 & 1.8 &  3.9  \\
5.0 & 2.5 & 0.1 && 5.0 & 1.7  & 0.07 & 1.1 & 4.7 && 4.0  & 3.5 & 0.48 & 1.8 & 3.7  && 4.1  & 3.3  & 0.49 & 1.8 & 3.7  \\
5.0 & 1.0 & 0.1 && 5.0 & 0.9  & 0.10 & 1.2 & 4.4 && 4.0  & 3.0 & 0.55 & 2.0 & 3.4  && 4.0  & 3.0  & 0.54 & 2.0 & 3.4  \\
\cline{1-3}\cline{5-9}\cline{11-15}\cline{17-21} \\
15.0 & 15.0 & 0.5 && 14.5 & 20.0 & 0.36 & 1.6 & 14.4 && 11.3 & 16.0 & 0.52 & 2.0 & 12.3 && 11.2 & 15.9 & 0.56 & 2.0 & 12.2  \\
15.0 & 10.0 & 0.5 && 13.4 & 12.5 & 0.39 & 1.8 & 12.1 && 10.3 & 11.7 & 0.52 & 2.0 & 10.3 && 10.2 & 11.9 & 0.56 & 2.0 & 10.3  \\
15.0 &  5.0 & 0.5 && 12.6 &  6.3 & 0.50 & 2.3 &  8.7 && 8.1 &  7.8 & 0.65 & 2.3 &  7.5 && 8.3 &  7.7 & 0.67 & 2.4 &  7.6  \\
10.0 & 10.0 & 0.5 && 10.0 &  11.5 & 0.35 & 1.6 & 9.8 &&  8.2 &  9.9 & 0.55 & 2.0 &  8.3 &&  8.1 &  9.8 & 0.57 & 2.0 &  8.2  \\
10.0 &  5.0 & 0.5 &&  8.9 &  5.2 & 0.39 & 1.8 &  7.0 &&  6.7 &  6.0 & 0.55 & 2.0 &  6.1 &&  6.8 &  5.9 & 0.57 & 2.1 &  6.1  \\
10.0 & 2.5 & 0.5 && 8.8 & 2.9  & 0.52 & 2.4 & 5.1 && 5.3  & 4.2 & 0.72 & 2.5 & 4.4  && 5.8  & 4.0  & 0.72 & 2.6 & 4.5  \\
 5.0 &  5.0 & 0.5 &&  5.3 &  4.6 & 0.35 & 1.7 &  5.0 &&  4.5 &  4.5 & 0.58 & 2.1 &  4.3 &&  4.5 &  4.5 & 0.60 & 2.1 &  4.3  \\
5.0 & 2.5 & 0.5 && 4.8 & 1.7  & 0.33 & 1.7 & 3.6 && 3.7  & 2.8 & 0.58 & 2.1 & 3.1  && 3.7  & 2.9  & 0.61 & 2.2 & 3.2  \\
5.0 & 1.0 & 0.5 && 4.7 & 0.9  & 0.50 & 2.1 & 2.5 && 3.5  & 1.6 & 0.77 & 2.7 & 2.1  && 3.6  & 1.6  & 0.77 & 2.8 & 2.1  \\
\cline{1-3}\cline{5-9}\cline{11-15}\cline{17-21} \\
15.0 & 15.0 & 0.9 && 13.6 & 19.4 & 0.63 & 2.3 & 14.8 && 12.2 & 17.4 & 0.62 & 2.2 & 13.6 && 11.8 & 17.3 & 0.68 & 2.4 & 13.7  \\
15.0 & 10.0 & 0.9 && 10.9 & 11.7 & 0.65 & 2.4 & 10.4 && 9.9 & 10.7 & 0.62 & 2.2 & 9.7 && 9.3 &  11.5 & 0.69 & 2.4 & 9.8  \\
15.0 &  5.0 & 0.9 &&  8.1 &  4.7 & 0.67 & 2.6 &  5.6 && 6.1 & 5.2 & 0.66 & 2.3 &  5.4 && 5.8 &  5.6 & 0.73 & 2.6 &  5.4  \\
10.0 & 10.0 & 0.9 && 9.9 & 11.5 & 0.63 & 2.3 & 9.9 && 9.3 & 10.5 & 0.63 & 2.2 &  9.3 && 9.0 &  10.6 & 0.68 & 2.4 &  9.3  \\
10.0 &  5.0 & 0.9 && 7.1 &  4.7 & 0.64 & 2.4 &  5.4 &&  5.9 &  5.0 & 0.64 & 2.2 &  5.2 &&  5.7 &  5.3 & 0.70 & 2.4 &  5.2  \\
10.0 & 2.5 & 0.9 && 5.3 & 2.1  & 0.73 & 2.7 & 2.9 && 3.7  & 2.4 & 0.70 & 2.4 & 2.8  && 3.7  & 2.5  & 0.74 & 2.6 & 2.8  \\
 5.0 &  5.0 & 0.9 &&  5.9 &  4.6 & 0.61 & 2.3 &  5.0 &&  5.4 & 4.7 & 0.64 & 2.3 &  4.8  &&  5.1 &  5.0 & 0.70 & 2.5 &  4.8  \\
5.0 & 2.5 & 0.9 && 4.4 & 1.7  & 0.60 & 2.3 & 2.7 && 3.5  & 2.1 & 0.63 & 2.2 & 2.7  && 3.5  & 2.3  & 0.70 & 2.4 & 2.7  \\
5.0 & 1.0 & 0.9 && 3.2 & 0.7  & 0.79 & 2.8 & 1.2 && 1.8  & 0.9 & 0.73 & 2.5 & 1.2  && 2.1  & 0.9  & 0.79 & 2.7 & 1.2  \\
\hline
\hline
\end{tabular}
\end{center}
\end{table*}

Figure~\ref{fig:fig12} presents the measured global structural parameters  $n_{\rm T}$ and $R_{\rm e,T}$ as functions of $B/T$ and the ratio of $R_{\rm e,B}$ to $R_{\rm e,D}$. We point out that the relationships between these parameters are similar for $R_{\rm e,D}$=5, 10 or 15\,pixels and we thus only show the results at $R_{\rm e,D}=10$\,pixels.
it is clear that the global \sersic index $n_{\rm T}$  is  not only tightly correlated with $B/T$ and but also affected by the sizes of disk and bulge components for individual bulge+disk models (the left panels).
We notice that the ratio $R_{\rm e,B}$/$R_{\rm e,D}$ has an effect on $n_{\rm T}$ because larger $R_{\rm e,B}/R_{\rm e,D}$ (relatively bigger bulge)  tends to lower $n_{\rm T}$ \citep[see also][]{Lang14}. On the other hand, the global effective radius $R_{\rm e,T}$  is sensitive to both the ratio $R_{\rm e,B}/R_{\rm e,D}$  and $B/T$, and approximately equals to the light-weighted combination of $R_{\rm e,B}$ and $R_{\rm e,D}$. When the bulge+disk models are dominated by one component (disk or bulge), the recovered structural parameters are close to these of the dominant component.
The recovered profiles (with PSF) are compared with the corresponding reference profiles (without PSF) to examine the PSF effect. The two sets of structural parameters agree well with each other, indicating that the global structural parameters of the composite-type galaxies can be properly recovered using the method of 1-D profile fitting as PSF effect has little influence on the recovery of the structural parameters.

The middle panels of Figure~\ref{fig:fig12} shows $n_{\rm T}$ and $R_{\rm e,T}$ recovered from the stacked profiles of bulge+disk models with spreads in $AR_{\rm D}$ and $R_{\rm e,D}$ included.  We can see that the recovered $n_{\rm T}$ and $R_{\rm e,T}$ exhibit significant deviations from those presented in the left panels at $B/T<0.5$; when the disk component increasingly dominates the total light, $n_{\rm T}$ is increasingly overestimated and $R_{\rm e,T}$ is more underestimated, and thus the recovered profile appears to be more compact, suggesting that it is the $AR$ and $R_{\rm e}$ spreads of the disk component causing the deviations in $n_{\rm T}$ and $R_{\rm e,T}$, in particular at $B/T<0.5$. These are consistent with the results in Section~\ref{sec:are}.
Similarly, the right panels of  Figure~\ref{fig:fig12} shows  $n_{\rm T}$ and $R_{\rm e,T}$ recovered from the stacked profiles of bulge+disk models accounting for spreads in $AR_{\rm D }$ and $R_{\rm e,D}$and $R_{\rm e,B}$ (right).   No significant difference is found from the middle panels, indicating that the bulge size spread $R_{\rm e,B}$ marginally affects the estimate of global structural parameters.

\subsection{Dual-\sersic Profile Fitting}\label{subsec:dual}

\begin{figure*}
\begin{center}
\includegraphics[width=0.75\textwidth]{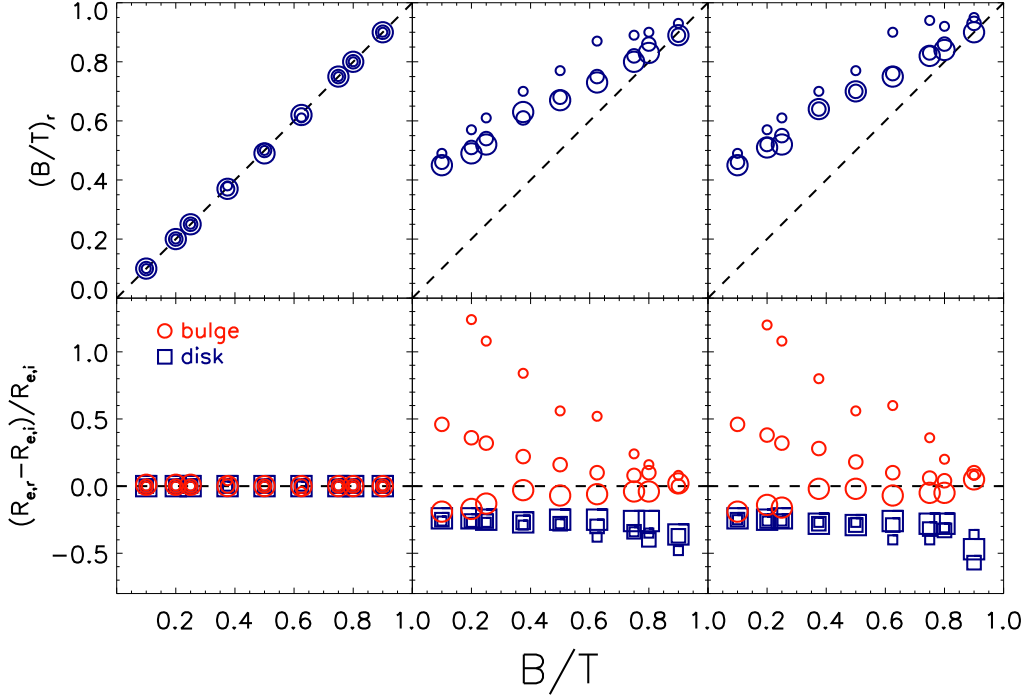}
\caption{Recovered $R_{\rm e}$ of the disk/bulge component ({\it bottom}) and $(B/T)_{\rm r}$ ({\it upper}) through the dual-\sersic profile fitting as a function of intrinsic $B/T$ for individual bulge+disk models ({\it left}), the stacked profiles of bulge+disk models with spreads in $AR_{\rm D }$ and $R_{\rm e,D}$ counted ({\it middle}), and the stacked profiles of bulge+disk models with spreads in $AR_{\rm D }$, $R_{\rm e,D}$ and $R_{\rm e,B}$ included ({\it right}). The deviation of the recovered $R_{\rm e}$ relative to the intrinsic value is shown in the bottom panels, and the deviation of the recovered $B/T$ relative to the intrinsic value is shown in the upper panels.}
\label{fig:fig13}
\end{center}
\end{figure*}

Resolving high-$z$ galaxies into bulge and disk components even in a statistical sense is key to drawing an empirical picture for bulge growth. Here we test the recovery of the structural parameters of the two components through fitting  the global 1-D profile of a stacked image with bulge+disk composite profiles.  We test this method and see how the measurements rely on the parameters of galaxy models for stacking.  In our two-component fitting, we fix \sersic index for the bulge ($n$=4) and the disk ($n$=1).  Again, the models presented in Table~\ref{tab:2com} are used in our simulations. The best fitting is selected using the least squares method across parameter space of $R_{\rm e,D}$, $R_{\rm e,B}$, and $B/T$. The best-fit dual-\sersic profile provides the recovered structural parameters $R_{\rm e,D}$, $R_{\rm e,B}$, and $B/T$ as the bulge and disk component of a target. The results of the dual-\sersic profile fitting are shown in Table~\ref{tab:2com}, in comparison with the input model parameters when $B/T$=0.1, 0.5 and 0.9.

The left panel of Figure~\ref{fig:fig13} shows the results of dual-\sersic fitting to the 1-D profile of individual bulge+disk models. with $R_{\rm e,D}=10$\,pixels and $R_{\rm e,B}$=[10, 5, 2.5]\,pixels.  Note that the results  are not dependent on $R_{\rm e,D}$.  One can see that the three key structural parameters $R_{\rm e,D}$, $R_{\rm e,B}$ and $B/T$ can be properly recovered using the method of dual-\sersic profile fitting.

Accounting the spreads of $AR$ and $R_{\rm e}$ for the disk component of bulge+disk models,  we derive the structural parameters of the bulge and disk component of the stacked image of the models through dual-\sersic profile fitting and show the results in the middle panel of Figure~\ref{fig:fig13}. Clearly, the $AR$ and $R_{\rm e}$ spreads  affect the recovery of the structural parameters of the two components.  While the average size of the disk component $<R_{\rm e,D}>$ is systematically underestimated by about 20\%, independent from $B/T$,  the average size of the bulge component $<R_{\rm e,B}>$ is increasingly overestimated at decreasing  $B/T$, particularly for lower $R_{\rm e,B}/R_{\rm e,D}$.  The  $B/T$ is increasingly overestimated at decreasing $B/T$. These biases in recovering the structural parameters are obviously caused by the inclined disks in stacking.    The degree of bias in $B/T$ is strongly dependent on the intrinsic $B/T$ but free from  $R_{\rm e,B}/R_{\rm e,D}$.


Along with the spreads of $AR$ and $R_{\rm,e}$ added to disks,  the spread in  bulge $R_{\rm,e}$ is also included in stacking of bulge+disk models. The right panel of Figure~\ref{fig:fig13} presents the recovered structural parameters using the method of dual-\sersic fitting to the global 1-D profile of the stacked model image. Again, the recovered structural parameters deviate from the input ones. The deviations of $<R_{\rm e,D}>$, $<R_{\rm e,B}>$ and $<B/T>$ are similar to what given in the middle panel of Figure~\ref{fig:fig13}, suggesting that the bulge $R_{\rm,e}$ spread has little influence on the estimate of the averaged structural parameters of the stacked bulge+disk models. And the biases in $<R_{\rm e,B}>$ and $<B/T>$ are mainly caused by the $AR$ and $R_{\rm e}$ spreads of the disk component.

\begin{figure*}
\begin{center}
\includegraphics[width=0.75\textwidth]{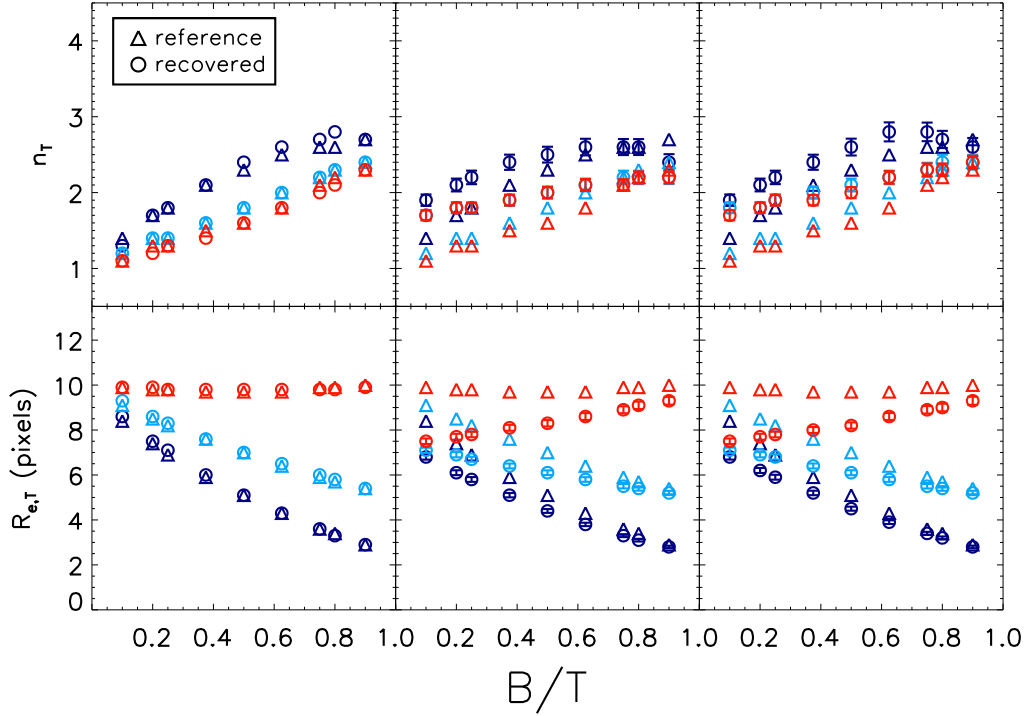}
\caption{Measured global structural parameters  $n_{\rm T}$ and $R_{\rm e,T}$  as functions of $B/T$ and the ratio of $R_{\rm e,B}$ to $R_{\rm e,D}$ for individual pseudo-bulge+disk models ({\it left}),  the stacked profiles of pseudo-bulge+disk models with $AR$ and $R_{\rm e}$ spreads added to the disk component ({\it middle}), and  the stacked profiles of pseudo-bulge+disk models with $AR$ and $R_{\rm e}$ spreads for the disk component and $R_{\rm e}$ spread for the bulge component  ({\it right}). Here $n=2.5$ is adopted for pseudo bulges and $n=1$ for disks.  Color codes $R_{\rm e,B}/R_{\rm e,D}$= 0.25 (blue), 0.5 (cyan) and 1 (red) at $R_{\rm e,D}$=10\,pixels. Circles represent the recovered profiles from PSF-convolved models, while  triangles in all panels mark the reference profiles from individual PSF-free models.}
\label{fig:fig14}
\end{center}
\end{figure*}

\subsection{Measurements of pseudo-bulges}

Not all bulges are classical type with a \sersic index of $n$=4. In fact, many bulges with smaller \sersic indices in the local universe are recognized as pseudo bulges and thought to be built up through secular evolution  \citep[e.g.,][and references therein]{Graham01,Balcells03,Kormendy04}.  For high-$z$ galaxies, the bulges formed through inward migration of disk clumps are expected to differ from the classical ones which are usually formed via mergers \citep[e.g.,][]{Bournaud07}.  Since the pseudo bulges are less distinct from disks compared to the classical bulges, it is important to examine how the global properties depend on the model parameters and to what degree the pseudo bulge component can be resolved through a dual-\sersic profile fitting to a 1-D surface brightness profile.  We therefore repeat the simulations in Section~\ref{subsec:two} but set $n$=2.5 for the bulge component.

Again,  we derive global structural parameters $n_{\rm T}$ and $R_{\rm e,T}$ by fitting single \sersic profiles to the simulated profiles in three cases: individual pseudo bulge+disk models, stacking of pseudo bulge+disk models with $AR$ and $R_{\rm e}$ spreads added to the disk component, and stacking of such models  with $R_{\rm e}$ spread added to the bulge component and $AR$ and $R_{\rm e}$ spreads added to the disk component. Table~\ref{tab:pseudo} lists the results only for $B/T=[0.1, 0.5, 0.9]$.
Figure~\ref{fig:fig14} shows recovery results of the global structural parameters. Similar to those present in Figure~\ref{fig:fig12}:  the recovered global $n_{\rm T}$ increases from 1 to 2.5 as $B/T$ increases from 0  to 1; when  $B/T$ is fixed,  a lower ratio $R_{\rm e,B}/R_{\rm e,D}$  generally leads $n_{\rm T}$ to be slightly higher; the global size  can be approximately seen as the light-weighted combination of $R_{\rm e,B}$ and $R_{\rm e,D}$; the recovered parameters agree well with the reference parameters ( triangles) derived from the PSF-free models; the  $AR$ and $R_{\rm e}$ spreads of the disk component leaves the global \sersic index  increasingly overestimated and the global size increasingly underestimated at decreasing $B/T$; the $R_{\rm e}$ spread of the pseudo bulge component has insignificant influence to the global structural parameters.

\begin{figure*}
\begin{center}
\includegraphics[width=0.75\textwidth]{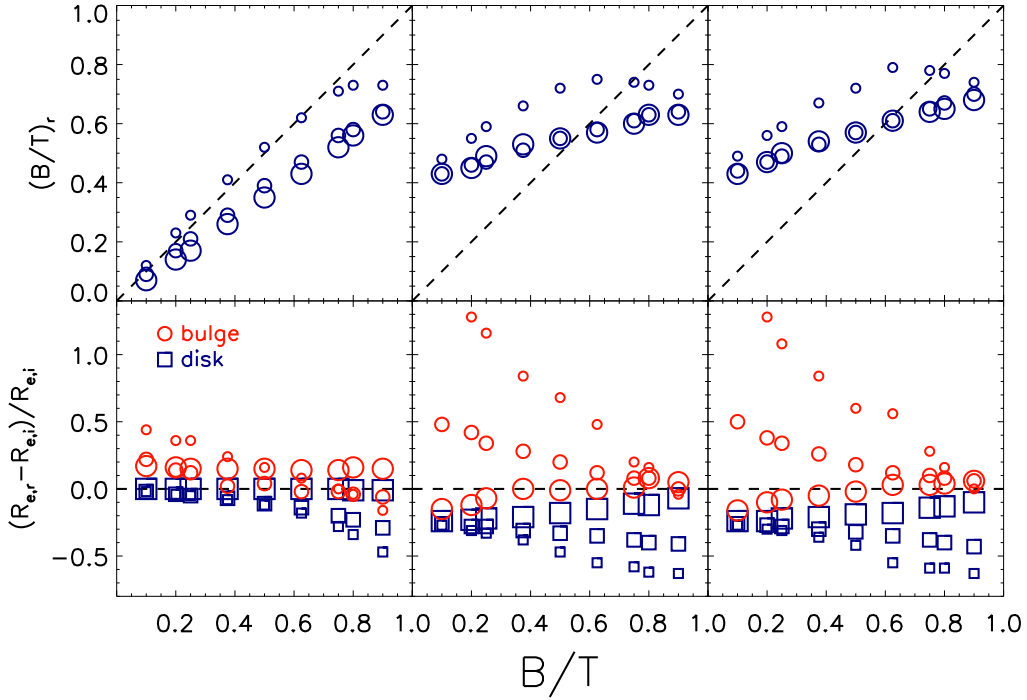}
\caption{Recovered  $R_{\rm e}$ ({\it bottom}) and $B/T$ ({\it upper}) as a function of $B/T$ through dual-\sersic profile fitting to the 1-D profiles of individual pseudo-bulge($n=2.5$)+disk models ({\it left}), the stacked profiles of pseudo-bulge+disk models with spreads in $AR_{\rm D}$ and $R_{\rm e,D}$ counted ({\it middle}), and the stacked profiles of pseudo-bulge+disk models with spreads in $AR_{\rm D}$, $R_{\rm e,D}$ and $R_{\rm e,B}$ included ({\it right}). The best-fit dual-\sersic profiles are taken as the bulge (red) and disk (blue) components. The deviations of $R_{\rm e}$ and $B/T$ relative to the input parameters are shown. in the fitting, the \sersic index of bulges is fixed at 4, which is not the case of the pseudo-bulge+disk models.}
\label{fig:fig15}
\end{center}
\end{figure*}

We point out that  fitting the stacked profiles of pseudo-bulge+disk models  with dual-\sersic profile results in similar results for classical bulge+disk models presented in Figure~\ref{fig:fig13} if the two \sersic profiles have $n=1$ for one (disk) and $n=2.5$ for the other (pseudo bulge).  Moreover, it is interesting to see how the recovery of structural parameters are affected when $n=4$ is adopted as the bulge component in the dual-\sersic profile fitting.
Accordingly, the recovered $R_{\rm e,B}$, $R_{\rm e,D}$ and $B/T$ as a function of input $B/T$ are presented in Figure~\ref{fig:fig15}, for individual pseudo bulge+disk models, stacking of these models with the spreads of  $AR$ and $R_{\rm e}$ included into the disk component, and stacking of these models with $AR$ spread counted for the disk component and $R_{\rm e}$ spread for both components, respectively.

When  the pseudo-bulge in the pseudo bulge+disk models is mistaken as a classical one, the bulge+disk decomposition ends up with  an increasing underestimate of $B/T$  (also $R_{\rm e,B}$ and $R_{\rm e,D}$) with increasing input $B/T$. This is understandable in that the pseudo bulge component  is divided into a classical bulge and an additional disk.  At $R_{\rm e,B}/R_{\rm e,D}<1$, the additional disk biases the recovered $R_{\rm e,D}$ smaller.
Taking the spreads of $AR$ and $R_{\rm e}$ into account, the recovery of the averaged structural parameters of $<R_{\rm e,B}>$, $<R_{\rm e,D}>$ and $<B/T>$ from the stacked profile of pseudo bulge+disk models is further affected mainly by  the $AR$ and $R_{\rm e}$ spreads of the disk, as described in Section~\ref{subsec:dual}.


\section{SUMMARY} \label{sec:disc}

Stacking technique is a powerful tool to probe signals under the detection limit of individual images for faint galaxies and enable us to obtain the averaged surface brightness profile toward larger radii.
We carried out simulations to test the stacking of galaxy models  with spreads in axis ratio ($AR$; elongation or inclination) and effective radius ($R_{\rm e}$) counted, and to explore how the recovered structural parameters of the averaged surface brightness profile depend on the effective radius ($R_{\rm e}$), axis ratio ($AR$), index of \sersic profile ($n$) of the models and their spreads. We also addressed the recovery of averaged structural parameters through stacking bulge+disk models in order to simulate the real observations.

In our simulations, we use circular apertures to derive surface brightness profile of a galaxy image. This is the usual way for stacking of faint high-$z$  galaxies, whose structural parameters are barely known so that corrections for inconsistence in inclination, orientation and size are often ignored. Following \cite{Szomoru12}, we fit the one-dimensional (1-D) radial surface brightness profile of a stacked galaxy image with a library of 1-D \sersic profiles to obtain the best-fit profile using the method of least squares. The best-fit profile is taken as the intrinsic profile for the stacked image. Galaxy models used in our simulations have structural parameters spanning sufficiently wide ranges: $1\le n \le 6$, $0\farcs 05 \le R_{\rm e} \le 0\farcs 75$ and $0.1 \le AR \le 1.0$.

We examined 1) the dependence of the measured surface brightness profile solely on each parameter of $n, R_{\rm e}$ and $AR$; 2) the recovery of structural parameters of the mean surface brightness profile when a set of galaxy models for stacking have two parameters fixed and the third parameter spreading within a certain distribution; 3) the recovery of structural parameters of the mean surface brightness profile of galaxy models with spreads counted in both $AR$ and $R_{\rm e}$ as functions of $n$ and   mean effective radius $<R_{\rm e}>$; 4) the recovery of structural parameters of the mean surface brightness profile of galaxy models with spreads added to all parameters, including $AR$, $R_{\rm e}$ and $n$,  for late-type and early-type galaxies; 5) the fitting of the stacked profile of bulge+disk models with both single \sersic profiles and  dual-\sersic profiles to see to which extent the stacking can recover the averaged structural parameters of a galaxy population.

The striking results from our simulations are that the structural parameters $n$ and $R_{\rm e}$ of the mean-stacked image of a group of galaxies may be biased up to 70\% by spreads in $AR$ and $R_{\rm e}$, much dependent on the distribution functions of the spreads; the bias can be quantitatively corrected once the spread functions are known.  We summarize our results as follows:

\begin{itemize}

\item The spread in $AR$ leads the mean-stacked image of a group galaxies to appear more compact, i.e., with a lower  mean effective radius $<R_{\rm e}>$ and a higher $n$.  The inclusion of highly-inclined (or elongated) galaxies with low $AR$ in stacks biases the recovered  $n$ and $<R_{\rm e}>$.

\item For early-type galaxies with large \sersic index $n$, which have extended halos in the outskirts, the oversubtraction of background biases the estimate of structural parameters. The mean \sersic index $n$ may be underestimated, and the effective radius $R_{\rm e}$ may be underestimated for early-type galaxies with large sizes. This indicates that the estimation of background is very important for the estimation of structural parameters, especially for early-type galaxies which have extended halos in the outskirts.

\item Accounting  for the $AR$  spread  of local disk galaxies from \cite{padilla08} in stacking galaxies with $n\leq 2.5$ , the median effective radius $<R_{\rm e}>$ of the stacked galaxies is underestimated by 23\% and the mean \sersic index $n$ is overestimated by up to 20\%.  Similarly, the $AR$ spread of local early-type galaxies from \cite{hao06} in stacking galaxies with $n>2.5$ leads to an underestimate of 12\% in $<R_{\rm e}>$ and little influence on the estimate of $n$.

\item The spread in $R_{\rm e}$ plays a different role from that of the spread in $AR$, which leads the mean-stacked profile of galaxies  to be more concentrated. Taking the log-normal distribution of $R_{\rm e}$  from \cite{shen03} into account, $<R_{\rm e}>$ is not biased for either early-type or late-type galaxies, and \sersic index $n$ is overestimated by 50\% for the late-type galaxies but not significantly biased for the early-type galaxies.
When  $R_{\rm e}$ scatters within a spread, the stacked galaxies with a size  smaller than the median contribute more to the central part of the integrated light, and those with a size larger than the median  contribute more to the extended wing, resulting in a mean-stacked profile with a higher \sersic index and an effective radius equal to the median of the $R_{\rm e}$ spread.

\item The effects of the spreads in $AR$ and $R_{\rm e}$ are linearly co-added on the estimate of structural parameters of a stacked image.  The actual corrections for these effects rely on the spread functions.

\item Account for the spread of $n$ with a uniform distribution, we find that the recovered structural parameters remain unchanged compared with these derived with no spread in $n$ counted, suggesting that the spread of $n$ does not significantly effect on the stacked results.

\item In stacking analysis, the galaxies are often classified by the stellar mass and type, and the $AR$, $R_{\rm e}$ and $n$ all have spreads. In this case, we find that the effective radius $R_{\rm e}$ can be underestimated by 20\% to 27\% for late-type galaxies, and only 10\% to 15\% for early-type galaxies, due to the $AR$ distribution of galaxies. The \sersic index $n$ can be well recovered for early-type galaxies, but can be overestimated for late-type galaxies due to both the $AR$ and $R_{\rm e}$ distribution.

\item For faint galaxies, the center we find will have an offset to the real center of galaxies due to the noise. We also test this effect by stacking galaxies by using the real center instead of the center found by Sextractor. We find that for galaxies in the range $24-24.75$\,mag, the 68\% percentile of the centering offset is not more than 1\,pixel and does not influence the results.

\item Recovery of structural parameters of bulge+disk galaxies are regulated by the bugle-to-total ratio $B/T$, the $R_{\rm e}$ ratio of the two components and the \sersic index of the bulge  if the disk is exponential  ($n=1$). The global  \sersic index $n_{\rm T}$ mainly depends on \sersic index of the bulge and $B/T$.  The global effective radius $R_{\rm e,T}$ can be approximately seen as the light-weighted combination of bulge size $R_{\rm e,B}$ and disk size$R_{\rm e,D}$. And the ratio $R_{\rm e,B}/R_{\rm e,D}$ also has influence on $n_{\rm T}$ at a second-order level.
The $AR$ and $R_{\rm e}$ spreads of the disks and the $R_{\rm e}$ spread of the bulges lead $n_{\rm T}$ to be increasingly overestimated and $R_{\rm e,T}$ to be increasingly underestimated at decreasing $B/T$.

\item By dual-\sersic profile fitting to the stacked profile of bulge+disk galaxies, the  averaged structural parameters of the bulge and disk components can be determined although the uncertainties are large.
The $AR$ and $R_{\rm e}$ spreads of disks is the main cause of the increasing overestimate of $B/T$ (also  $R_{\rm e,B}$) at decreasing intrinsic $B/T$. The issue of the bulges to be pseudo or classical would dramatically change the results of the dual-\sersic profile fitting for a composite profile.  We caveat that the measurement of bulge growth in a statistical sense is significantly affected by the $AR$  and $R_{\rm e}$ spreads of the disks and \sersic index of the bulges in the bulge+disk decomposition through stacking. We thus stress that interpretation of stacking results should take these biases into account and the corrections for the biases are strongly dependent on the structural parameters of stacked galaxies and the spread functions of them.

\end{itemize}

\acknowledgments

We are grateful to the  referee for the valuable suggestions,  which significantly improved this manuscript.
This work is supported by by the Strategic Priority Research Program "The Emergence of Cosmological Structures" of the Chinese Academy of Sciences (Grant No. XDB09000000), National Basic Research Program of China (973 Program 2013CB834900) and NSFC grant (U1331110).

\end{document}